\documentclass[review]{elsarticle}

\journal{Journal of Sound and Vibration}

\usepackage[ngerman,english]{babel}  
\usepackage[german]{nomencl}
\usepackage[utf8]{inputenc}

\usepackage{footnote}
\makesavenoteenv{tabular}
\makesavenoteenv{table}

\usepackage{microtype}

\usepackage{url}
\usepackage{eurosym}    
\usepackage{subfigure}  
\usepackage{geometry}   
\usepackage{amsmath}    
\usepackage{amssymb}
\usepackage{graphicx}  
\usepackage{epstopdf}
\usepackage{multicol}
\usepackage{multirow}
\usepackage{placeins}
\usepackage{bm}
\usepackage{setspace}  
\usepackage{tabularx}
\usepackage{wallpaper}
\usepackage[absolute]{textpos}
\usepackage{float}
\usepackage[bottom]{footmisc}
\usepackage{booktabs}  
\usepackage{caption}

\usepackage{units}     
\usepackage{nicefrac}  

\usepackage{pdfpages} 

\usepackage{listings}                                   
\usepackage[multidot]{grffile} 

\usepackage[T1]{fontenc}         
\usepackage{tikz}           
\usetikzlibrary{patterns}
\usepackage{epsfig}

\usepackage{listings}    

\usepackage{xcolor}
\usepackage{soul}

\usepackage[bookmarksopen = true, linkcolor = black, plainpages = false,hypertexnames = false,citecolor = black]{hyperref}
\definecolor{dkgreen}{rgb}{0,0.6,0}    
\definecolor{gray}{rgb}{0.5,0.5,0.5}
\definecolor{mauve}{rgb}{0.58,0,0.82}

\usepackage{xargs}          
\usepackage[colorinlistoftodos,prependcaption,textsize=tiny]{todonotes}
\newcommandx{\unsure}[2][1=]{\todo[linecolor=red,backgroundcolor=red!25,bordercolor=red,#1]{#2}}
\newcommandx{\change}[2][1=]{\todo[linecolor=blue,backgroundcolor=blue!25,bordercolor=blue,#1]{#2}}
\newcommandx{\info}[2][1=]{\todo[linecolor=dkgreen,backgroundcolor=dkgreen!25,bordercolor=dkgreen,#1]{#2}}
\newcommandx{\improve}[2][1=]{\todo[linecolor=mauve,backgroundcolor=mauve!25,bordercolor=mauve,#1]{#2}}
\newcommandx{\thiswillnotshow}[2][1=]{\todo[disable,#1]{#2}}

\newlength{\mylen}                  
\newcommand{\vlen}{\the\mylen}      

\renewcommand{\div}{\nabla\cdot}

\newcommand{\deltalam}{\delta}

\renewcommand{\eta}{\mu}

\newcommand{\yplus}{y^{+}}

\newcommand{\kappav}{\kappa_{\mathrm{v}}}

\newcommand{\Uz}{w}

\newcommand{\vecu}{\bm{u}}

\newcommand{\tensL}{\mathbf{L}}

\newcommand{\rhoO}{\rho_0}

\newcommand{\coh}{\gamma_{i}^2(f)}
\newcommand{\csd}{G_{q_\textrm{a,ref}q_{\textrm{a,}i}}(f)}
\newcommand{\psdx}{G_{q_\textrm{a,ref}q_\textrm{a,ref}}(f)}
\newcommand{\psdy}{G_{q_{\textrm{a,}i}q_{\textrm{a,}i}}(f)}

\newcommand{\cO}{c} 

\begin{document}

\begin{frontmatter}

\title{Numerical investigation of a deep cavity with an overhanging lip considering aeroacoustic feedback mechanism}

\cortext[mycorrespondingauthor]{Corresponding author}
\author{S. Schoder\fnref{myfootnote}\corref{mycorrespondingauthor}}
\ead{stefan.schoder@tuwien.ac.at}

\author{I. Lazarov\fnref{myfootnote}\corref{}}

\author{M. Kaltenbacher\fnref{myfootnote}\corref{}}

\fntext[myfootnote]{Institute of Mechanics and Mechatronics, Vienna University of Technology, 1060 Vienna, Austria}

\begin{abstract}
In modern transport systems, passengers' comfort is greatly influenced by flow-induced noise. 
In this study we investigate a generic deep cavity with an overhanging lip, mimicking a door gap in a vehicle, that is overflowed by air at two different free stream velocities, $\unit[26.8]{m/s}$ and $\unit[50]{m/s}$.
The turbulent boundary layer and the acoustic waves interact with the cavity's geometry and form a strong feedback mechanism.
In the present work, we focus on the details of the compressible turbulent flow structures and their variations concerning the velocity, the boundary layer as well as the domain dimensionality for a later acoustic simulation within a hybrid aeroacoustic workflow.
Furthermore, we verify the feasibility of reducing the acoustic computational domain from 3D to 2D for this application by conducting a coherence study of acoustically active flow structures in the spanwise direction.
The role of the three-dimensional Taylor-G{\"o}rtler vortices from the recirculation regarding the vortex formation and the vortex-edge interaction was also evaluated.
Remarkably for the lower approaching velocity ($\unit[26.8]{mm}$), we found a special vortex-edge interaction, namely an alternating sequence of complete clipping and a subsequent partial escape.
%
%
Lastly, we assigned previous unknown peaks in the pressure spectrum to their corresponding mechanisms.

\end{abstract}


\end{frontmatter}



\section{Introduction}

In modern transport systems, passengers' comfort is greatly influenced by flow-induced noise. A cavity with a lip represents a generic model of a vehicle door gap, involving an acoustic feedback mechanism on the underlying flow field.
Due to complex interactions between acoustic, vortex, and entropy modes (see~\cite{chu58}), various sound classifications of cavity problems have been introduced over the years.
In the late 1970s, Rockwell and Naudascher~\cite{rockwell78} proposed a classification into three mechanisms: fluid-dynamic, fluid-resonant, and fluid-elastic.
Fluid-dynamic modes, known as Rossiter modes, are related to an aerodynamic feedback mechanism, which energetically feeds self-sustaining oscillations.
On the other hand, fluid-resonant modes arise from acoustic resonance, e.g. Helmholtz resonance. 
%
%
This paper considers these two mechanisms and neglects fluid-elastic interaction, which represents a fluid-structure coupling resulting from oscillations of the cavity walls.
If a fluid-dynamic and a fluid-resonant mode coincide, they can combine in the so-called lock-on state.
In case of deep rectangular cavities ($D>L$) at low Mach numbers $M<0.18$, it is likely that one of the first two Rossiter modes lock with the cavity's depth mode~\protect\cite{east66}.

\begin{figure}[h!]
    \centering
        \subfigure[Cavity with microphone (black dot) positioned in the middle of the leading wall.]{\includegraphics[width=0.34\textwidth]{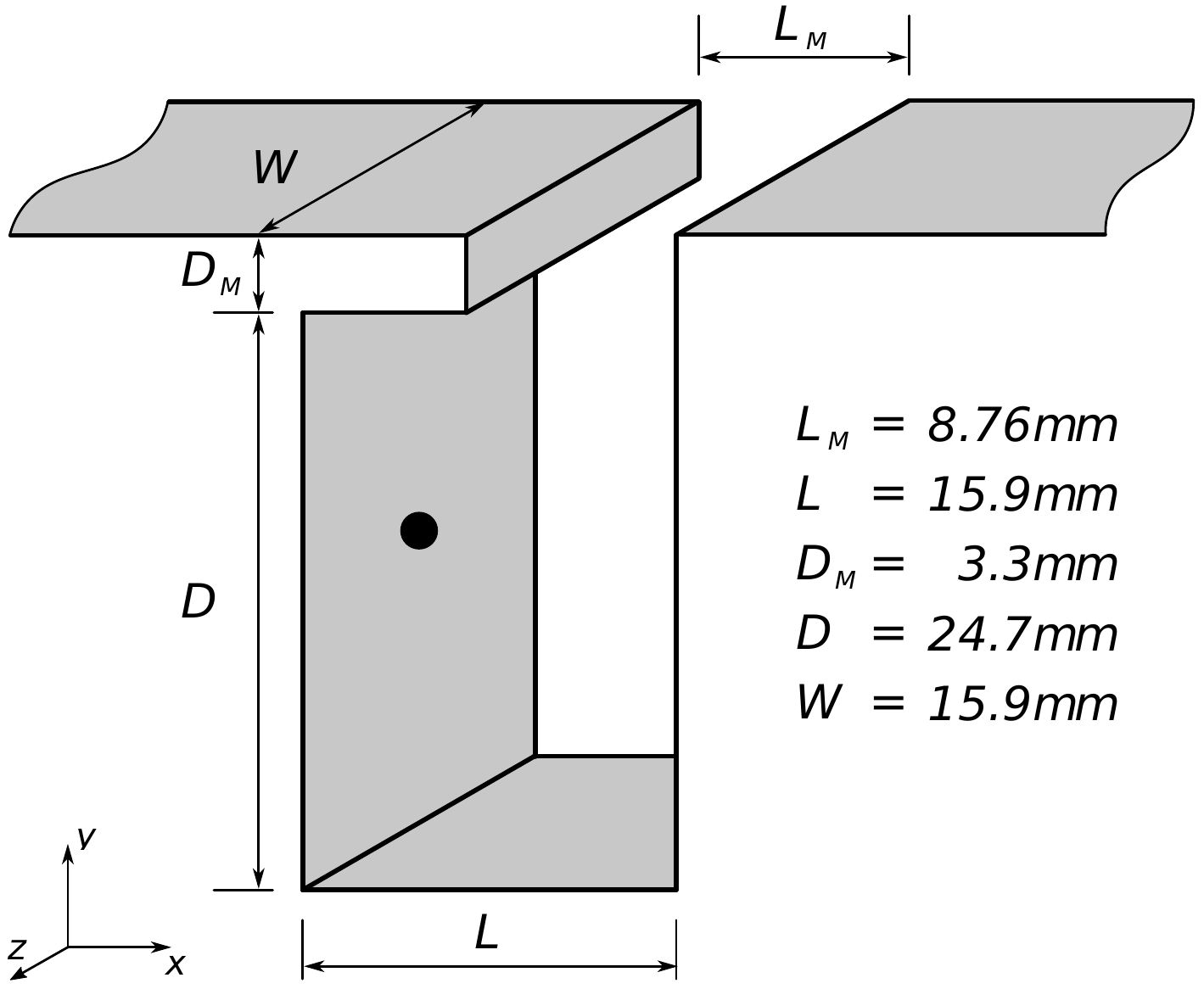}\label{fig:cavity_sketch}}    
        \hfill
        \subfigure[Simulation domain in $\unit{mm}$ and prescribed boundary conditions.]{\includegraphics[width=0.64\textwidth]{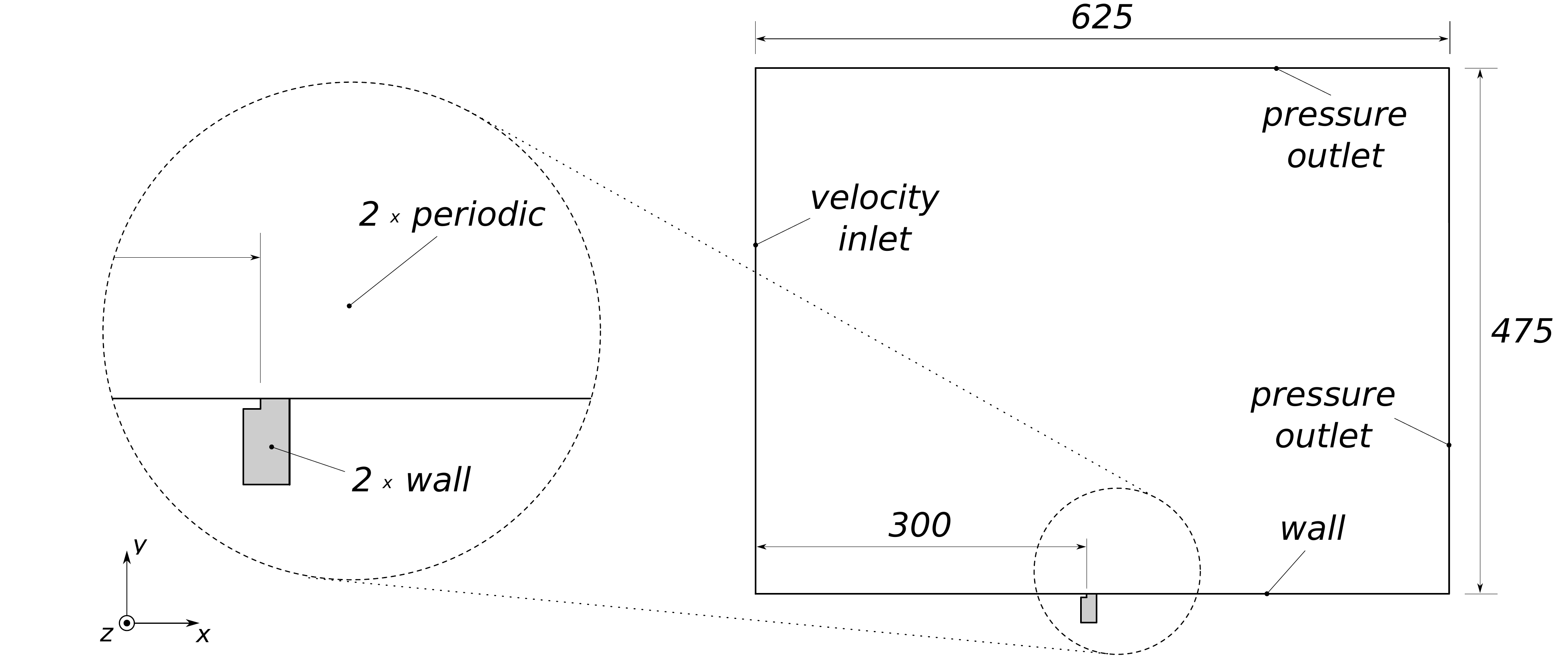}\label{fig:cfd_domain_drawing}}
    \caption{Cavity geometry and computational domain.}
    \label{fig:cavity_sketch_and_domain}
\end{figure}

%
We investigate a generic deep cavity with an overhanging lip that is overflowed by air at two different free stream velocities, namely $\unit[26.8]{m/s}$ and $\unit[50]{m/s}$. The turbulent boundary layer and the acoustic waves interact with the cavity's geometry and form a strong feedback mechanism.
Figure~\ref{fig:cavity_sketch} illustrates the geometry and the problem definition that were initially introduced and experimentally studied within the \textit{Third Computational Aeroacoustics (CAA) Workshop on Benchmark Problems} by NASA~\cite{nasa3CAABenchmark2000} and Henderson~\cite{henderson2000}, respectively.
Motivated by previous numerical studies (e.g.~\cite{farkasANDpaal2015,ahujaUndMendoza95}), we altered the CAA benchmark geometry from the prescribed cavity width~$W = \unit[150]{mm}$ to $W = \unit[15.9]{mm}$. According to Ahuja and Mendoza~\cite{ahujaUndMendoza95}, this change has an insignificant influence on the emitted sound, as long as the cavity width~$W$ exceeds the cavity mouth length~$L_\mathrm{M}$. Investigations on the coherence of acoustically active structures confirm this reduction. Consequently, we lowered the number of finite volume cells and thus the computational burden~\cite{lazarov2018}. 
%
%
While the flow velocities have been well documented in the experimental study~\cite{henderson2000}, the documentation of the boundary layer thicknesses were contradictory. As proposed by Farkas and Paal~\cite{farkasANDpaal2015}, we use a boundary layer thickness of $\unit[10]{mm}$ at the cavity's leading edge (reference simulation). Furthermore, we vary the boundary layer thickness from $\unit[6-14]{mm}$ in our parametric study. In the experimental study, the fluid-dynamic pressure, including the sound pressure, is recorded inside the cavity.

This cavity problem has been numerically investigated in 2D (see~\cite{moon2000,kurbatskii2000,ashcroft2000,lin2004,zhang2004,loh2004}) during the third and the fourth CAA Workshop on Benchmark Problems~\cite{nasa3CAABenchmark2000,nasa4CAABenchmark2004}.
%
%
%
Zhang et al.~\cite{zhang2004} summarized the conducted numerical studies concerning Henderson's~\cite{henderson2000} experimental results.
The authors showed that despite the different nature of the flow fields (incompressible or compressible) and the used numerical approaches (URANS and DNS), the workshop participants have succeeded in qualitatively reproducing the experimental data.
In most numerical studies the arising frequencies were underestimated and the pressure levels overestimated compared to the experiments.
%
%
Zhang et al.~\cite{zhang2004} relate these frequency deviations to a thinner boundary layer that is used in most simulations.
Besides, Lin et al.~\cite{lin2004} assume that the pressure level deviations are a consequence of the incompletely resolved turbulent structures and the used turbulence models.
The authors pointed out
that discrete tones are related to vortex detachment, whereas broadband sound components are related to turbulence.
In general, throughout both CAA workshops it was found that the numerical results are strongly dependent on various parameters, such as the flow velocity, boundary layer thickness, grid refinement (especially in the vicinity of the cavity orifice), time step size, as well as dimensions (2D/3D) and size of the computational domain.
Unwanted acoustic reflections at the domain boundaries impose an additional increase in both tone frequency and pressure level and lead to erroneous results (see~\cite{kurbatskii2000,ashcroft2000}).
Therefore, a characteristic acoustic radiation boundary should be imposed on the acoustic free field boundaries and too small computational domains should be avoided.
According to Kurbatskii and Tam~\cite{kurbatskii2000}, at least one wavelength should be used as a rule of thumb for this purpose.

Over the years, further 2D studies reinvestigated this cavity 
~\cite{ashcroft2003,koh2003,wang2007}.
Carrying out a 2D compressible URANS study, Ashcroft et al.~\cite{ashcroft2003} have shown that the oscillation frequencies of the Rossiter mode and the pressure levels are proportional to the flow velocity.
Using a hybrid method, the authors have quantified the radiation directivity patterns of this cavity as monopole in the far field. In addition, an inverse relationship between the boundary layer thickness and the semi-empirical constant~$\kappav$ of the Rossiter's formula
\begin{equation}
    f_{\mathrm{R}_n} = \frac{U}{L_{\mathrm{M}}} \frac{n-\alpha}{M+\kappav^{-1}}, \text{~~~~~~~}  n\in \mathbb{N^+} \, ,
    \label{eq:RossiterGL}
\end{equation}
which represents the ratio between the vortex convection speed and the flow velocity, has been observed.
In Equ.~\eqref{eq:RossiterGL}, $f_{\mathrm{R}_n}$ denotes the $n^{\mathrm{th}}$ Rossiter mode, $U$ the free stream velocity, $L_{\mathrm{M}}$ the length of the cavity mouth, $M$ the Mach number, and $\alpha$ the time delay between the moment of the vortex impinging on the trailing edge and the emission of the acoustic waves.

The latest publication of Farkas and Paal~\cite{farkasANDpaal2015} studies the previous findings by suitable simulations and model variations.
For these purposes, the authors have investigated the influence of various turbulence models (both in 2D and 3D domains) and flow parameters.
Despite the relatively low Mach numbers of approximately $0.077$ and $0.144$ (where compressible effects are negligible compared to vortical effects), the authors have found that the compressible and incompressible fields differ significantly.
This behavior is in agreement with the numerical studies from Wang et al.~\cite{wang2007}.
While in case of a compressible fluid the flow oscillates in the $1^{\mathrm{st}}$ Rossiter mode (corresponds to one vortex in the cavity mouth), the incompressible fluid oscillates in the $2^{\mathrm{nd}}$ Rossiter mode.
%
Within a short parametric study Farkas and Paal~\cite{farkasANDpaal2015} showed that a change in the fluid viscosity causes no significant influence on the shear layer oscillation frequency (similar to~\cite{ahujaUndMendoza95}).
Furthermore, the authors had their biggest difficulties in dealing with acoustic reflections from the boundaries of the computational domain.
Thus, it is highly recommended to use suitable non-reflecting boundary conditions at the free field boundaries.

In contrast to~\cite{farkasANDpaal2015}, the present paper focuses on the details of the compressible turbulent flow structures and their variations concerning the velocity, the boundary layer as well as the domain dimensionality for a later acoustic simulation within a hybrid aeroacoustic workflow. Since incompressible flow simulations lead to insufficient results, we exclusively use a compressible fluid model to gain a profound understanding of the cavity.
The compressible flow equations are solved using finite volume methods as provided by ANSYS~Fluent~18.0~\cite{ansysTheory}. Large turbulent scales are resolved by a DES turbulence model on a 3D domain and in the far-field, we treat the acoustic component with acoustically absorbing boundaries based on the radiation characteristics. Furthermore, the grid study quantifies the most appropriate domain size and grid density for the numerical simulation. Overall, this work presents a robust setup for the flow simulation that may be used in a hybrid aeroacoustic method. As already mentioned, we verify the feasibility of reducing the acoustic computational domain from 3D to 2D for this application by conducting a coherence study of acoustically active flow structures in the spanwise direction. Thus we aim to quantify the influence of the 3D flow effects that originate from the recirculation flow and are attenuated by the cavity's sidewalls. To our best knowledge, three-dimensional effects, such as Taylor-G\"ortler vortices have not been presented before, and so the flow field and its acoustically active structures are investigated in the spanwise direction.
Remarkably for the lower approaching velocity ($\unit[26.8]{mm}$), we found a special vortex-edge interaction, namely an alternating sequence of complete clipping and a subsequent partial escape, which is in agreement with the experimental results of Rockwell and Knisely~\cite{rockwellANDknisely80} for cavities without an overhanging lip.

The rest of this paper is organized as follows: In Sec. \ref{chap:numericalSetUp}, we present the simulation setup and the grid convergence study. Section \ref{sec:parameterstudie_CFD} discusses parametric variations of the free stream velocity, the boundary layer thickness, and the effect of a reduced timestep size. Afterward, the 3D structures and the domain reduction for acoustic simulations are illustrated. The results of all simulations are then discussed in Sec. \ref{sec:fazitUNDausblick} and discrete pressure peaks are labeled by a source mechanism. Finally, Sec. \ref{sec:end} concludes our findings.

\section{Simulation setup}
\label{chap:numericalSetUp}

We consider the compressible fluid dynamics equations using air, modeled as an ideal gas at ambient conditions ($p=\unit[101325]{Pa}$ and $T=\unit[300]{K}$). The partial differential equations are solved by a pressure-based solver and second-order spatial and temporal schemes as provided by ANSYS Fluent 18.0 \cite{ansysTheory}.
The geometry of the investigated cavity is sketched in Fig.~\ref{fig:cavity_sketch}, and Fig.~\ref{fig:cfd_domain_drawing} depicts a side-view of the computational domain as well as the boundary conditions. All walls, including the cavity's spanwise side walls, are modeled as perfectly smooth, non-penetrating, no-slip walls. The rest of the spanwise domain boundaries are periodic (see Fig.~\ref{fig:cfd_domain_drawing}). At the top and the outlet, a pressure outlet combined with a non-reflecting boundary condition for the compressible waves is introduced. These non-reflecting boundaries are based on the characteristics of the Euler equation. At the inlet, we prescribe the boundary layer profile accounting for the boundary layer thickness at the cavity and again a non-reflecting boundary condition. The boundary layer profiles are obtained by an auxiliary stationary flat plate simulation.

\begin{figure}[h!]
    \centering
        \includegraphics[width=1.00\textwidth]{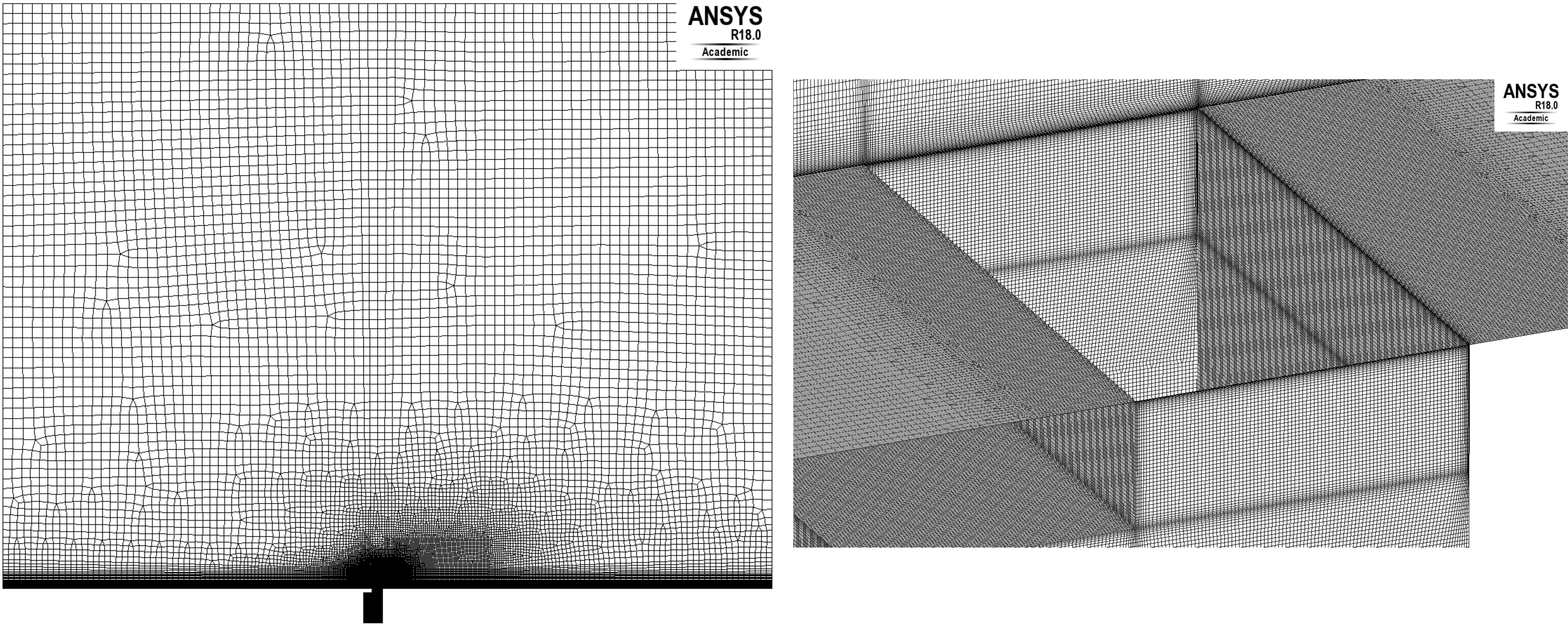}
    \caption{2D (left) and 3D (right) view of the fine computational grid consisting of approximately $11.3$ million cells.}
    \label{fig:CFDmesh_screenshot}
\end{figure}

\begin{figure}[t!]
    \centering
        \includegraphics[width=1.00\textwidth]{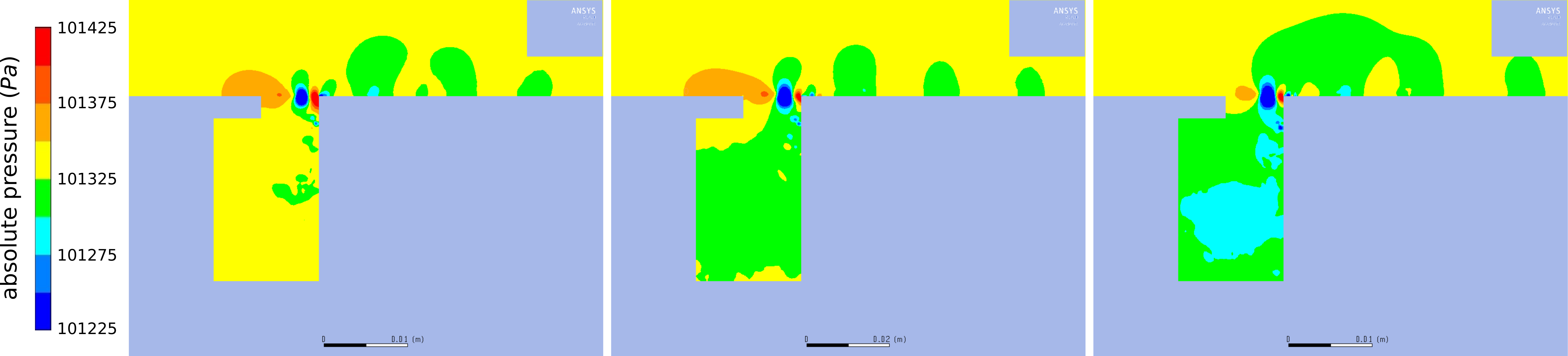}
    \caption{Slice in the middle plane (at $z=\unit[0]{mm}$) of the instantaneous pressure field for G-C (left), the G-M (middle) and G-F (right).}
    \label{fig:konvStud-pFelder-SBES}
\end{figure}

\begin{figure}[b!]
    \centering
        \includegraphics[width=0.80\textwidth]{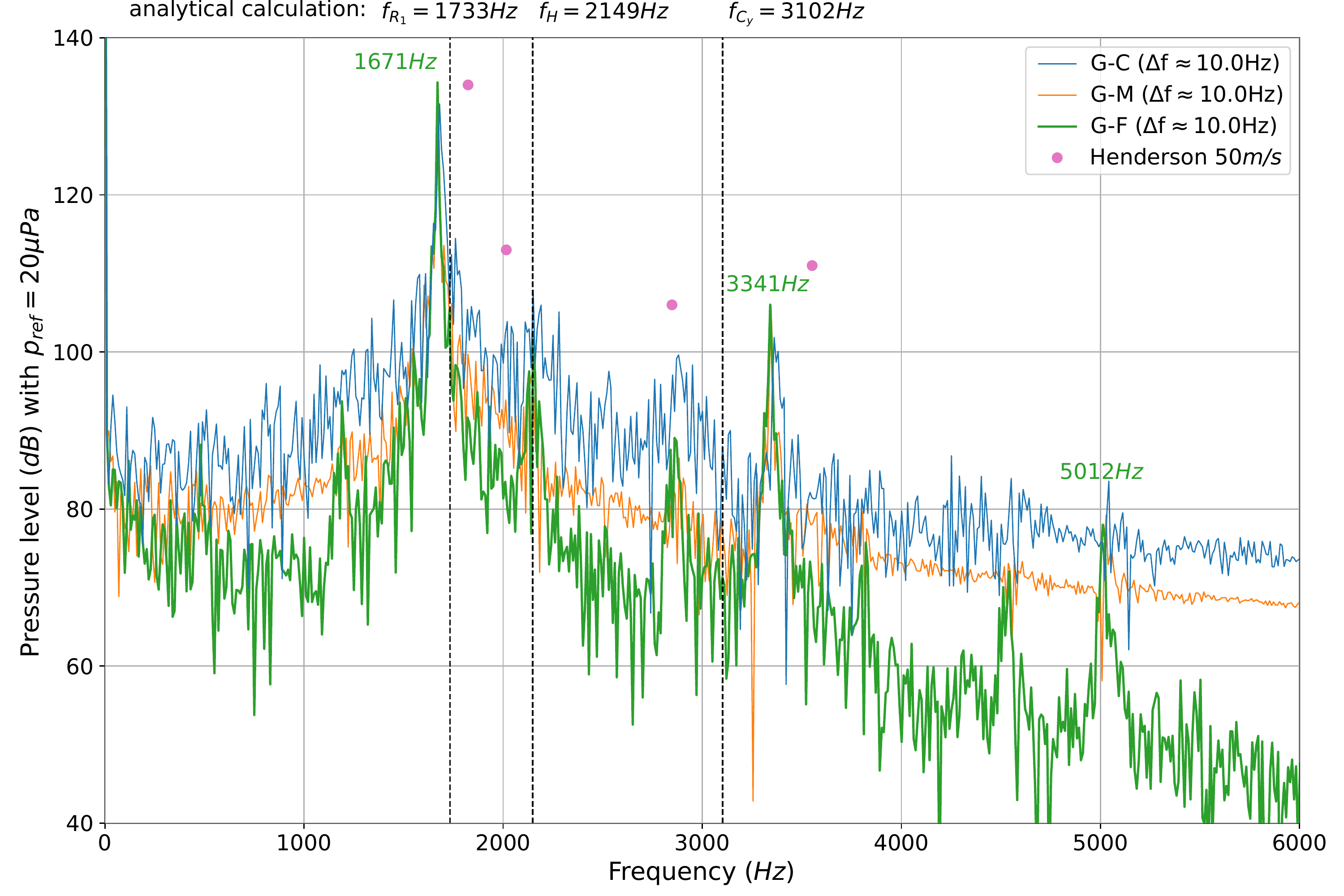}
        \caption{Pressure level spectra for the convergence study compared to the experimental data of Henderson~\cite{henderson2000}.
        }
    \label{fig:konvStud_FFTresults_normalTime_SBES}
\end{figure}

Although a structured grid could be easily generated for this simple geometry, we used a hybrid multi-block grid to reduce the number of finite volume cells. Each block is connected by identical discretizations at the boundaries, whereas different discretizations are used inside the blocks. Smooth and conform grid coarsening connects the different discretization densities inside these blocks. In this sense, the grid gradually becomes coarser with increasing distance from the cavity mouth.
Figure~\ref{fig:CFDmesh_screenshot} shows the finite volume grid that is designed for the flow velocity of $\unit[50]{m/s}$ and the SBES (Stress-Blended Eddy Simulation) turbulence model, that is, a DES (Detached Eddy Simulation) type turbulence model. This grid consists of approximately $11.3$ million cells and is denoted as the fine grid in this paper. The maximum length of the cell edges inside the cavity volume is approximate $\unit[0.1]{mm}$, whereas the maximum cell length outside of the cavity is $\unit[8]{mm}$. Especially in the vicinity of the cavity's mouth and at the walls, free and wall-bounded shear layers need to be properly resolved. With the presented setup, a $\yplus$ value of $0.97$ at the leading edge of the cavity can be achieved. To do so, $46$ wedge cells with a cell height of the first cells of $\unit[0.014]{mm}$ and a growth ratio of $1.1$ are used. This simulation setup considers the expected parameter variations throughout the parametric study in the next section.
Additionally to velocity scale based grid preparation, a grid convergence study verifies our simulation setup and determines the grid dependency of the results. The discretizations used for the grid convergence study are obtained by doubling the initial cell volumes~$\Delta V$ of the fine grid. 
In contrast to URANS (Unsteady Reynolds-Averaged Navier-Stokes) turbulence models, where only the $1^{\mathrm{st}}$ Rossiter mode and its higher harmonics are captured, the SBES model resolves more turbulent structures and we assess the broad-banded pressure spectrum at the microphone position in our grid study. The instantaneous pressure fields (see Fig.~\ref{fig:konvStud-pFelder-SBES}) and the corresponding pressure spectra (see Fig.~\ref{fig:konvStud_FFTresults_normalTime_SBES}) reveal that the flow oscillates in the $1^{\mathrm{st}}$ Rossiter mode. Compared to
the coarse (G-C) and the middle grids (G-M), the fine grid (G-F) resolves more turbulent flow structures. Hence we focus the discussion on the fine grid and chose it as a reference case for the following parametric study. Qualitatively, the three pressure spectra computed on different grids resolve similar structures (see Fig.~\ref{fig:konvStud_FFTresults_normalTime_SBES}). Besides the dominant $1^{\mathrm{st}}$ Rossiter mode ($\unit[1671]{Hz}$) and its higher harmonics ($\unit[3341]{Hz}$ and $\unit[5012]{Hz}$), further acoustic resonant modes are visible. The amplitude of the $1^{\mathrm{st}}$ Rossiter mode ($\unit[134.3]{dB}$) is well reproduced compared to the experimental data of Henderson~\cite{henderson2000} ($\unit[134]{dB}$).
In agreement with previous numerical studies, the arising frequency is underestimated with a relative deviation of approximately $\unit[8.39]{\%}$.
This discrepancy could be explained by the differences in the boundary layer thickness used in our simulation and those presented in the measurements. As shown in our parametric study, the peak frequencies and pressure levels are inversely dependent on the thickness of the approaching boundary layer above the leading edge of the cavity. Furthermore, the peak around $\unit[3552]{Hz}$ does not appear in Henderson's~\cite{henderson2000} discussion. According to our simulation, this is the $1^{\mathrm{st}}$ harmonic of the $1^{\mathrm{st}}$ Rossiter mode. Although the literature concerning this cavity problem uses the term sound pressure level for describing the pressure spectra, the correct term is pressure level, since the pressure signal obtained at the microphone position (see Fig.~\ref{fig:cavity_sketch}) includes the overall pressure (not just the acoustic part).

We assign the peak at $\unit[2151]{Hz}$ to the expected Helmholtz resonance frequency
\begin{equation}
    f_\mathrm{H}  = \frac{c}{2\pi} \sqrt{\frac{\pi R'^2}{V(D_\mathrm{M}+\frac{\pi R'}{2})}} = \unit[2149]{Hz},
    \label{eq:helmholzGL}
\end{equation}
calculated with the speed of sound~$c$ ($\unit[347.411]{m/s}$), the total cavity volume~$V$, the depth of the cavity mouth~$D_\mathrm{M}$, and the equivalent hydraulic radius~$R'=\sqrt{A/\pi}$, where $A$ stands for the area of the cavity orifice.
This resonant peak is comparable to the one from Henderson~\cite{henderson2000} ($\unit[2016]{Hz}$) and Loh et al.~\cite{loh2004} ($\unit[2062]{Hz}$).
%
Henderson~\cite{henderson2000} has no explanation for the peak around $\unit[2861]{Hz}$. We assume this acoustic resonance is a transversal cavity duct mode in the depth direction 
%
%
\begin{equation}
    f_\mathrm{C_y} = \frac{c}{4\left(D+D_\mathrm{M}\right)} =\unit[3102]{Hz}
    \label{eq:ductMode}
\end{equation}
modulated by the cavity orifice, where $c$ denotes the speed of sound and $\left(D+D_\mathrm{M}\right)$ the total cavity depth.
However, the other lowest longitudinal cavity resonances are outside the range of investigation at about $\unit[11]{kHz}$.
The peak at $\unit[1190]{Hz}$ in the pressure spectrum of the simulation using the fine grid may be a result of recirculation or vortex pairing. According to Loh et al.~\cite{loh2004}, this peak could be a subharmonic of the Helmholtz resonance.
Nevertheless, further studies are needed to classify the origin of this peak.

Additionally to the physical peaks, we detected an artificial computational domain resonance at $\unit[480]{Hz}$ of $\unit[88.2]{dB}$ in the pressure spectrum of the simulation on the fine grid. This non-physical resonance arises due to not fully absorbing boundary conditions and acoustic wavelength coincidences with the computational domain size. For the coarse and medium grid density, this peak is masked by the turbulent fluctuations at the measurement location.

To conclude the grid convergence study, the most appropriate computational grid is the one with the fine discretization. In the following section, we investigate the influence of various flow velocities~$U$, boundary layer thicknesses~$\deltalam$, and time-step sizes~$\Delta t$ for the fine grid.


\section{Parametric study}
\label{sec:parameterstudie_CFD}
Table~\ref{tab:CFDSimulationen-Ueberblick} summarizes the performed CFD simulations during the grid and the parametric studies.
Based on the the fine grid (G-F), parameter changes are highlighted by bold symbols. In total, we performed nine CFD simulations with different parameter combinations and computational grids.
\begin{table}[h!]
 \begin{center}
    \begin{tabular}{lcccccc}
        \toprule
        
        \textbf{Simulation code} & \textbf{Grid} & $\boldsymbol{\deltalam}$ & $\boldsymbol{U}$ & $\boldsymbol{\Delta t}$ & \textbf{Steps}  & $\boldsymbol{h_{\mathrm{CPU}}}$\\
        &  & $(\unit{mm})$ & $(\unit{m/s})$ & $(\unit{\mu s})$ & $(\unit{-})$  & $(\unit{h})$\\
        
        \midrule
          Grid study    & \\          
          G-C    & \textbf{coarse}   & $9.68$ & $50$ & $20$ & $7500$           & $1282$ \\
          G-M    & \textbf{middle}   & $9.68$ & $50$ & $20$ & $7500$           & $2808$ \\
          G-F    & \textbf{fine}     & $9.68$ & $50$ & $20$ & $7500$  & $6247$\\ 
          
          \midrule
          Parameter study &  \\
          P-D06  & fine & $\boldsymbol{6.06}$  & $50$   & $20$ & $7500$       & $6366$ \\
          P-D08  & fine & $\boldsymbol{8.06}$  & $50$   & $20$ & $7500$       & $5652$ \\
          P-D12  & fine & $\boldsymbol{11.96}$ & $50$   & $20$ & $7500$       & $5440$ \\
          P-D14  & fine & $\boldsymbol{14.03}$ & $50$   & $20$ & $7500$       & $6000$ \\ 
          P-U26  & fine & $9.7$   & $\boldsymbol{26.8}$ & $20$ & $10000$      & $6814$ \\
          P-T06  & fine & $9.68$  & $50$   & $\boldsymbol{1}$ & $114000$ & $49532$\\ 
          
          \bottomrule
         \end{tabular}
          \caption{CFD simulations during the parametric and the grid convergence study.
          The specified processor hours~$h_{\mathrm{CPU}} \approx N_{\mathrm{CPU}} \cdot t_{\mathrm{run}}$ are estimated by the number of processors~$N_{\mathrm{CPU}}$ and the simulation run time~$t_{\mathrm{run}}$. Note that the step numbers presented here are overall time steps. For data evaluation we used the last time span of approximately $\unit[1]{s}$, which corresponds to $\unit[5000]{}$ steps in case of $\Delta t = \unit[20]{\mu s}$ and $\unit[100000]{}$ steps for $\Delta t = \unit[1]{\mu s}$.}
           \label{tab:CFDSimulationen-Ueberblick}
 \end{center}
\end{table}

\subsection{Boundary layer thickness}
\begin{figure}[ht!]
    \centering
        \includegraphics[width=1.0\textwidth]{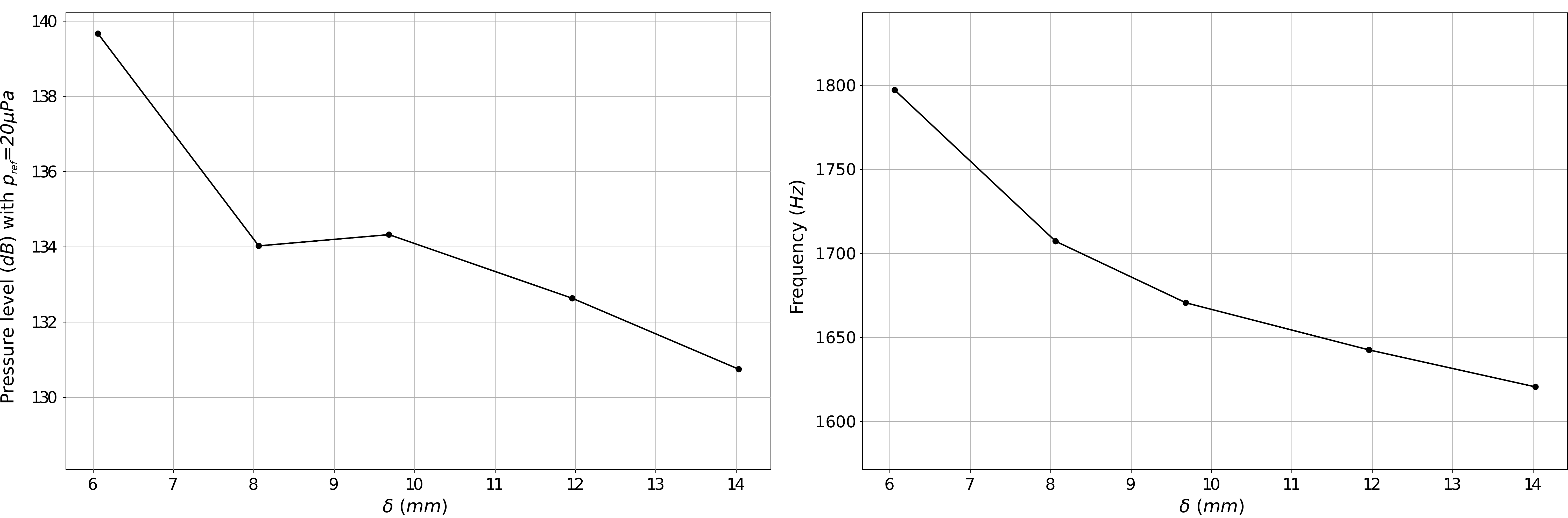}
    \caption{Pressure level (left) of the $1^{\mathrm{st}}$ Rossiter peak and its corresponding frequency~$f$ (right) as a function of the boundary layer thickness~$\deltalam$.}
    \label{fig:parameterstudie_delta_trends_f_SPL}
\end{figure}

Figure~\ref{fig:parameterstudie_delta_trends_f_SPL} shows the pressure level of the  $1^{\mathrm{st}}$ Rossiter mode and its peak frequency as a function of the boundary layer thickness~$\deltalam$. Both quantities exhibit an inverse proportional monotonic decrease for the boundary layer thickness~$\deltalam$ (e.g. see ~\cite{farkasANDpaal2015}).
This behavior meets the expectations that a thinner boundary shear layer potentially excites stronger oscillations and increases the pressure level. Familiar with the fundamentals of Rossiter's formula, the increasing frequency is a consequence of a higher convective speed of disturbances inside thinner shear layers. A comparison of the result of Rossiter's formula to the flow resonance suggests that our analytically used convection speed ($\kappav=0.43$) is the appropriate one for a boundary layer thickness of $\delta = \unit[8.06]{mm}$. Furthermore, in literature, both phenomenons are described by the mass reduction of a thinner shear layer leading to higher-frequency oscillations. 

Interestingly, a jump in the pressure level of roughly $\unit[3]{dB}$ occurs for the simulation with $\delta = \unit[8.06]{mm}$. Henderson~\cite{henderson2000} addressed this switching phenomena as a random process. 
%
%
%
In contrast to the monotonous character of the pressure level, previous 2D URANS or viscous flow simulations have shown non-monotonous (see~\cite{kurbatskii2000,ashcroft2003}) or even constant behavior (see~\cite{zhang2004}). According to the aforementioned literature, these deviations can be explained by the turbulence models in 2D, combined with the viscous flow, the stability of the main flow, and the boundary conditions. The authors explained their findings partly by the stability characteristics of the main flow profile. Concerning the present 3D study, we assume that the dimensionality of the computational flow domain allows vortex paring as well as recirculation and consequently a change in the shear layer dynamics.
Furthermore, Zhang et al.~\cite{zhang2004} showed for 2D simulations that different boundary conditions on the top boundary change the pressure level amplitude. For the pressure outlet boundary condition,  the pressure level amplitude remained constant for a varying boundary layer thickness. In contrast to this, the symmetry boundary condition leads to a pressure level drop with increasing boundary layer thickness. In this case, the pressure outlet condition with a characteristic boundary for far-field radiation is the appropriate choice.
In addition to the thickness, the shape of the prescribed boundary layer differs widely throughout the studies. Initially, the benchmark case was proposed with a one-seventh power law for the boundary layer. Since this definition does not represent reality, we focused our study on a developed turbulent boundary layer on a plate.

\subsection{Flow velocity}
\begin{figure}[ht!]
    \centering
        \includegraphics[width=0.8\textwidth]{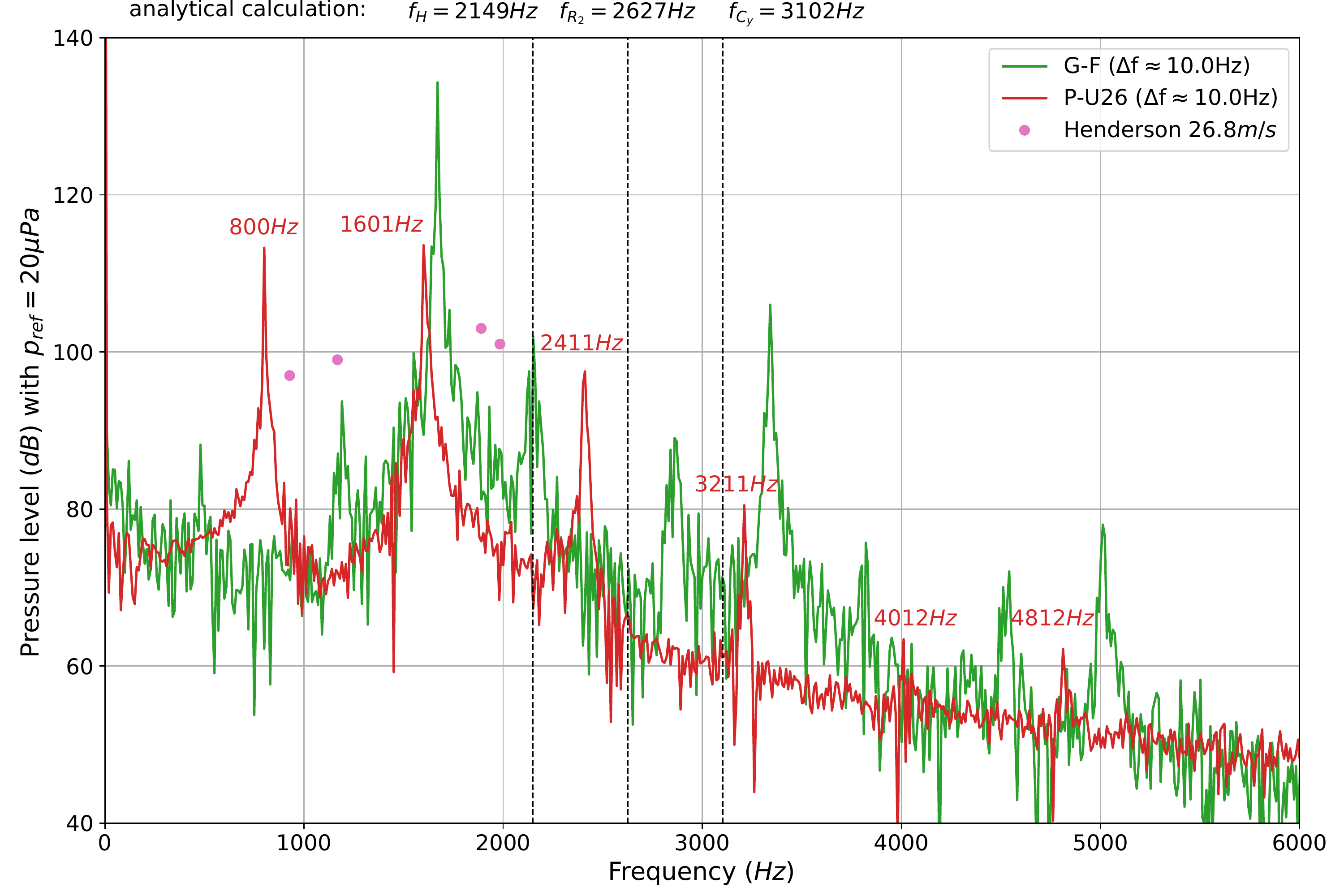}
    \caption{Pressure level spectra for the flow speed variation of the parametric study compared to the experimental data of Henderson~\cite{henderson2000}.
    }
    \label{fig:parameterstudie_speed_FFT}
\end{figure}

Figure~\ref{fig:parameterstudie_speed_FFT} shows the influence of the velocity variation on the pressure level fluctuations. Similar to~\cite{farkasANDpaal2015}, our study deviates from Henderson's experiments~\cite{henderson2000} at a first glance. We found that a partial vortex-edge interaction causes a subharmonic peak at $\unit[800]{Hz}$. 
Farkas and Paal~\cite{farkasANDpaal2015} accounted for their discrepancy to the low approach velocity, at which the driving mechanisms of the fluid-dynamic and fluid-resonant modes are more competitive than in the case of higher flow velocities.
During our investigation, the number of discrete peaks decreases for lower flow velocities, which we attribute to lower turbulent kinetic energy inside the flow.

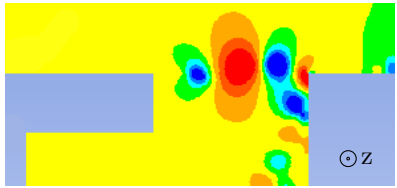
\begin{figure}[ht!]
    \centering
        \begin{tikzpicture}
            \node [inner sep=0pt, text width=0.32\textwidth]
            {\includegraphics[width=\textwidth, trim = 5.5cm 15.5cm 18cm 5cm, clip]{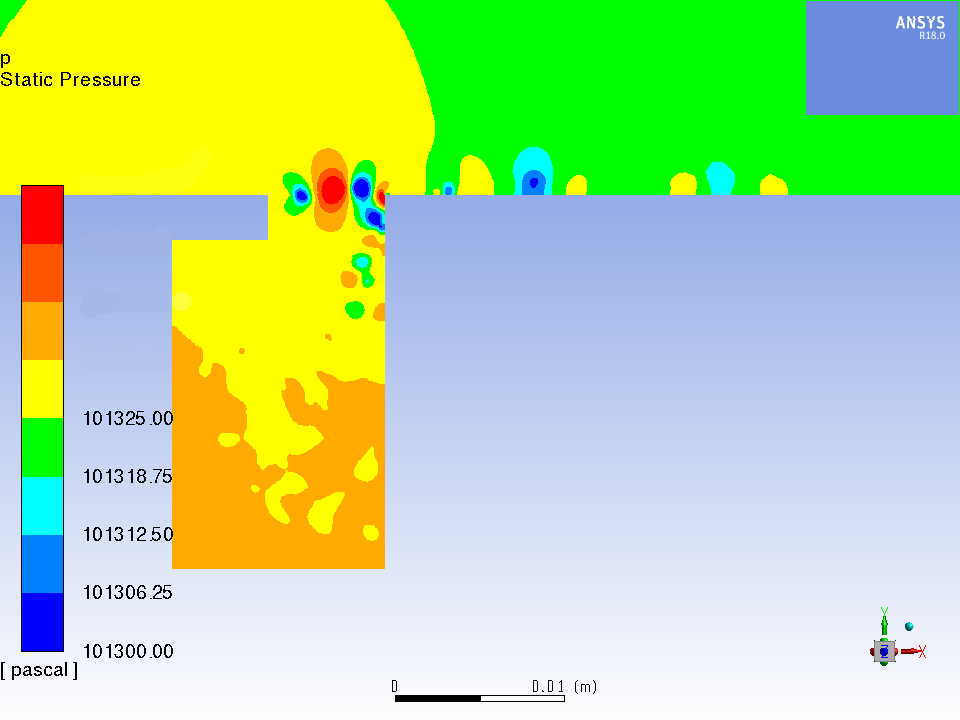}};
            \draw[draw=black, thin]  ( 1.95,-0.85) circle (0.5pt);
            \draw[draw=black, thin]  ( 1.95,-0.85) circle (3pt);
            \node (Z) at (2.20,-0.85) {z};  
        \end{tikzpicture}
    \caption{Slice in the middle plane (at $z=\unit[0]{mm}$) of the instantaneous static pressure~$p_{\mathrm{stat}}$ field for P-U26.
    Scaling of the $p_{\mathrm{stat}}$ values between $\unit[101300]{Pa}$ (blue) and $\unit[101350]{Pa}$ (red).
    }
    \label{fig:parameterstudie_speed_contour}
\end{figure}

Henderson~\cite{henderson2000} addresses the pressure level peak at $\unit[1168]{Hz}$ to a fluid-dynamic mode that analytically corresponds to the $1^{\mathrm{st}}$ Rossiter mode. Observing the pressure field (see Fig.~\ref{fig:parameterstudie_speed_contour}), we indicate that the flow does not oscillate in the $1^{\mathrm{st}}$ but in the $2^{\mathrm{nd}}$ Rossiter mode, which correlates to~\cite{farkasANDpaal2015}. We assume that the change in the Rossiter mode was misinterpreted by Henderson's experiment through the occurrence of the subharmonic peak. 
The already mentioned vortex edge interaction and the different boundary layer thickness explains the difference in the frequency and the amplitudes. Concluding from the higher flow speeds, doubling the boundary layer thickness can reduce the resonance peaks by almost $\unit[10]{dB}$, as long as the flow structures remain unchanged.
In our case, the air trapped inside the cavity oscillates with two expansion and two compression phases per period (see~\cite{koh2003}).
Our findings raise the question if the indicated peaks from Henderson~\cite{henderson2000} may be reassigned to a different source mechanism.
To clarify these findings, a further study focusing on the lower flow velocity ($\unit[26.8]{m/s}$) should be conducted.

A profound coherence study in the spanwise direction indicated that the recirculating flow inside the cavity plays an important role for both the P-U26 and the G-F cases (see Tab.~\ref{tab:CFDSimulationen-Ueberblick}).
While the flow structures around the shear layer were mainly two-dimensional, three-dimensional effects (Taylor-G{\"o}rtler vortices) inside the cavity mouth participate in the main vortex formation inside the shear layer. Figure~\ref{fig:SRC_coherence_explanation} shows that small-scale vortices shed from the lower edge of the cavity lip, driven by the three-dimensional recirculating flow, and interact with the shear layer instability by pushing in the vertical direction.
In this manner, an alternating sequence of complete clipping and a subsequent partial escape vortex-edge interaction (see~\cite{rockwellANDknisely80}) can be observed (see Fig.~\ref{fig:zweiteRossiterMode_SPU26_Uz}).
This means that only every second vortex hits the trailing edge of the cavity while the other vortex partially escapes the cavity.

\begin{figure}[ht!]
    \centering
        \begin{tikzpicture}
            \node [inner sep=0pt, text width=0.32\textwidth]
            {\includegraphics[width=\textwidth, trim = 5.5cm 14.5cm 18cm 6cm, clip]{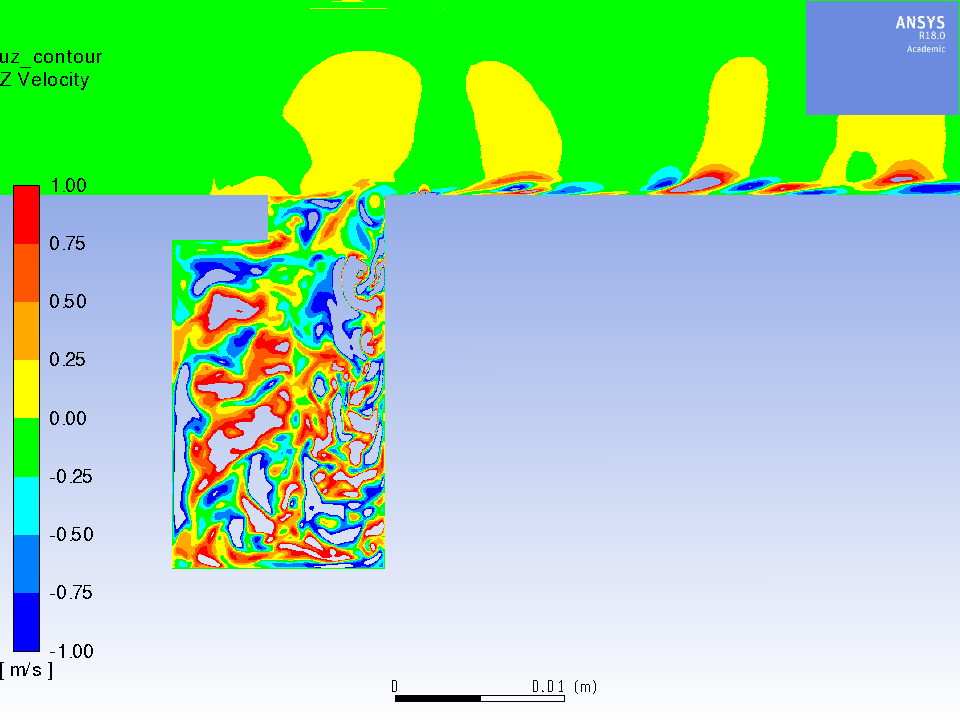}};
            \draw[draw=black, thick] (-0.05, 0.30) circle (10pt);
            \draw[draw=black, thin]  ( 1.95,-0.85) circle (0.5pt);
            \draw[draw=black, thin]  ( 1.95,-0.85) circle (3pt);
            \node (A) at (0.00, 0.45) {};   
            \node (B) at (0.50, 1.25) {};   
            \node (Z) at (2.20,-0.85) {z};  
            \draw[->, very thick, dashed, to path={-| (\tikztotarget)}]
              (A) -- (B);
            \end{tikzpicture}
        \includegraphics[width=0.32\textwidth, trim = 5.5cm 14.5cm 18cm 6cm, clip]{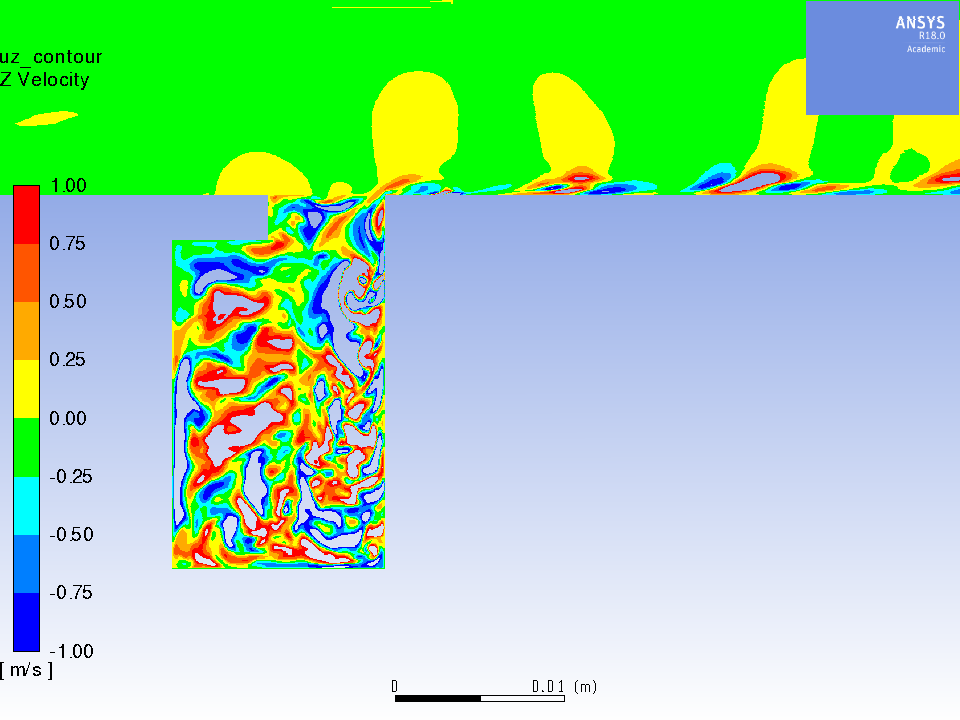}
        \includegraphics[width=0.32\textwidth, trim = 5.5cm 14.5cm 18cm 6cm, clip]{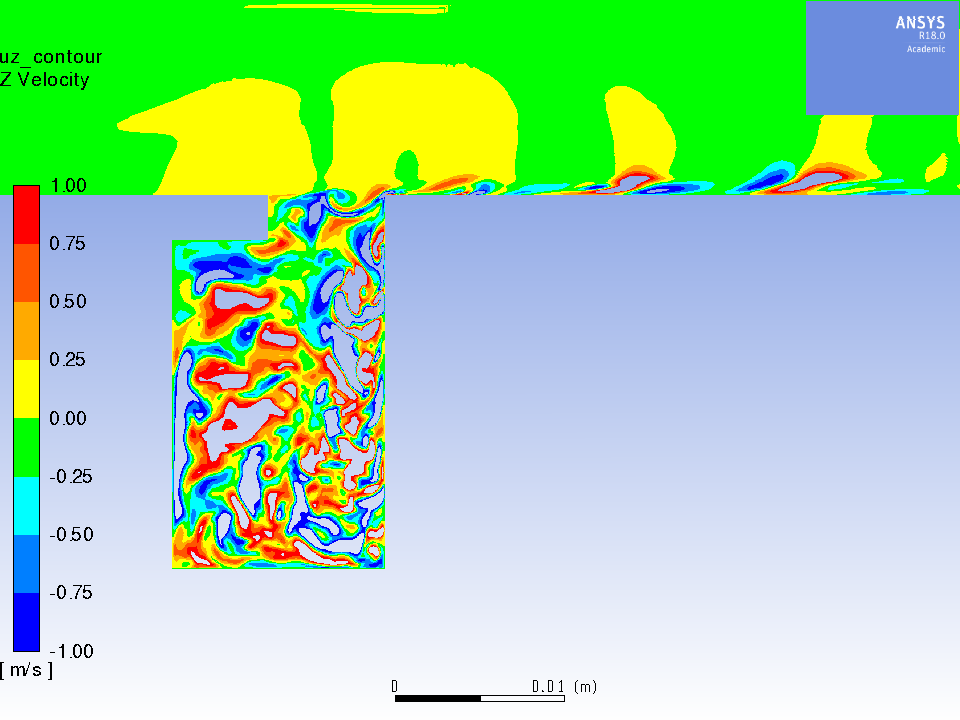}\\
        \vspace{0.5mm}
        \includegraphics[width=0.32\textwidth, trim = 5.5cm 14.5cm 18cm 6cm, clip]{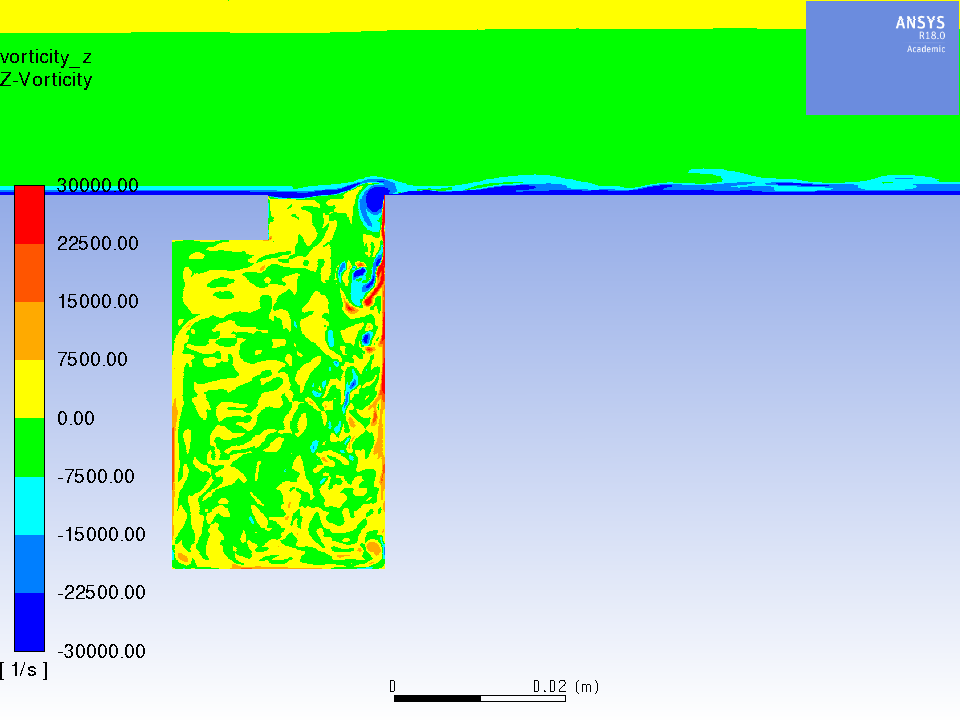}
        \includegraphics[width=0.32\textwidth, trim = 5.5cm 14.5cm 18cm 6cm, clip]{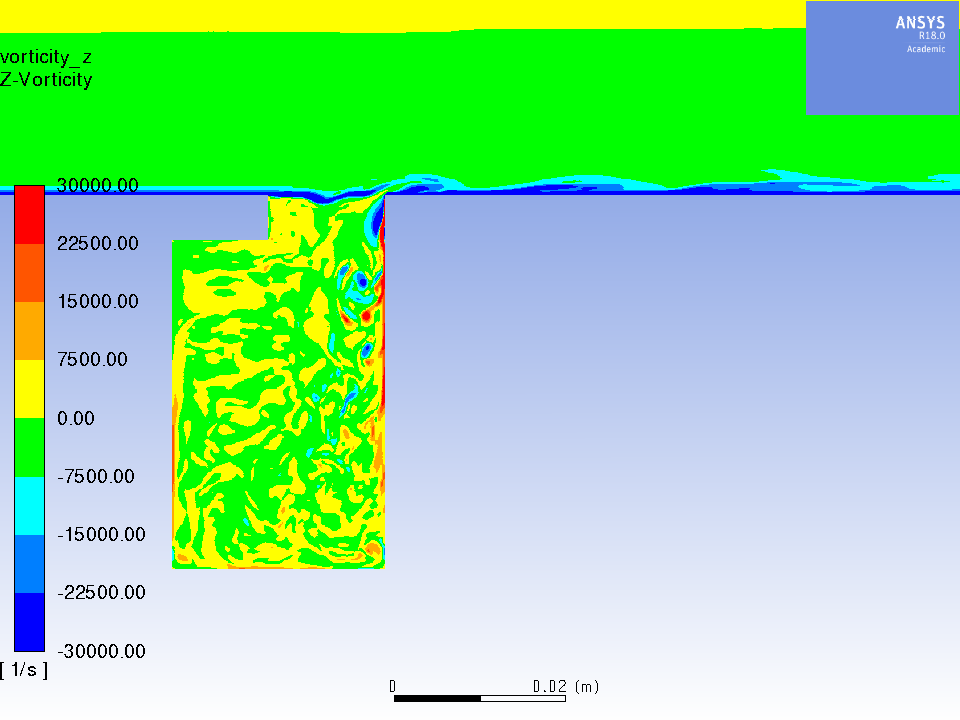}
        \includegraphics[width=0.32\textwidth, trim = 5.5cm 14.5cm 18cm 6cm, clip]{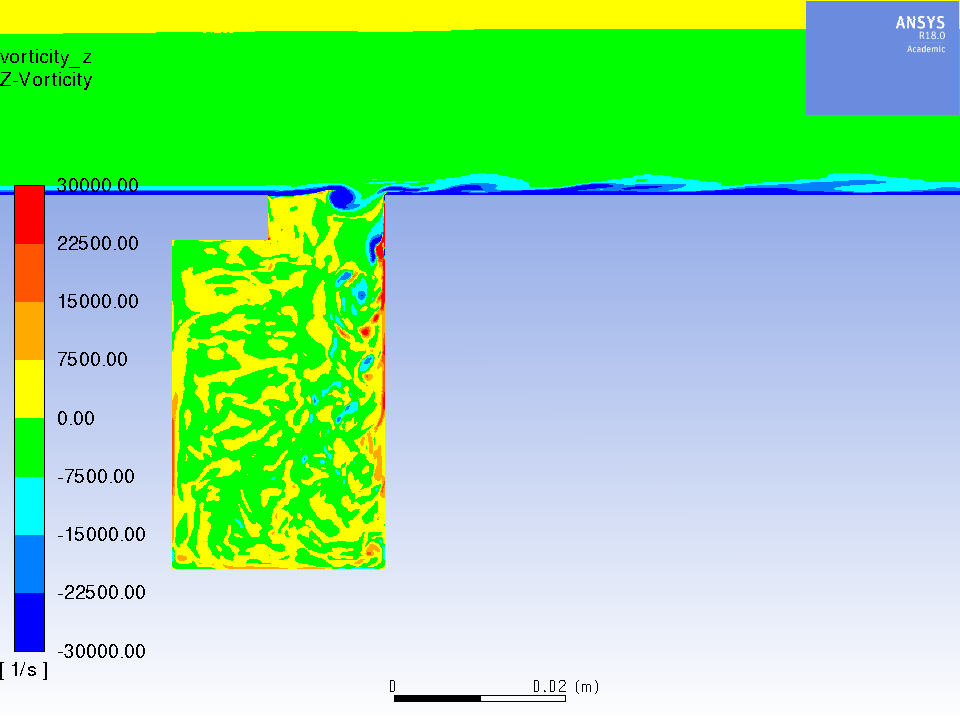}
    \caption{Slice in the middle plane displaying the instantaneous flow velocity~$\Uz$ field (top) and the vorticity~$\Omega_z$ (bottom) from G-F. The snapshot sequence for one time period is illustrated from left to the right. $\Omega_z$ scales between $\unit[-3\cdot 10^{4}]{1/s}$ (blue) and $\unit[3\cdot 10^{4}]{1/s}$ (red). $\Uz$ scales between $\unit[-1]{m/s}$ (blue) and $\unit[1]{m/s}$ (red).}
    \label{fig:SRC_coherence_explanation}
\end{figure}
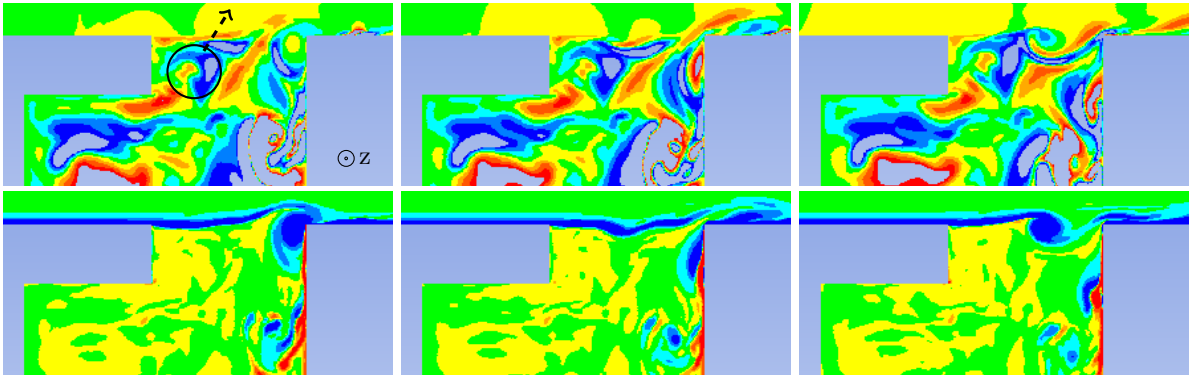
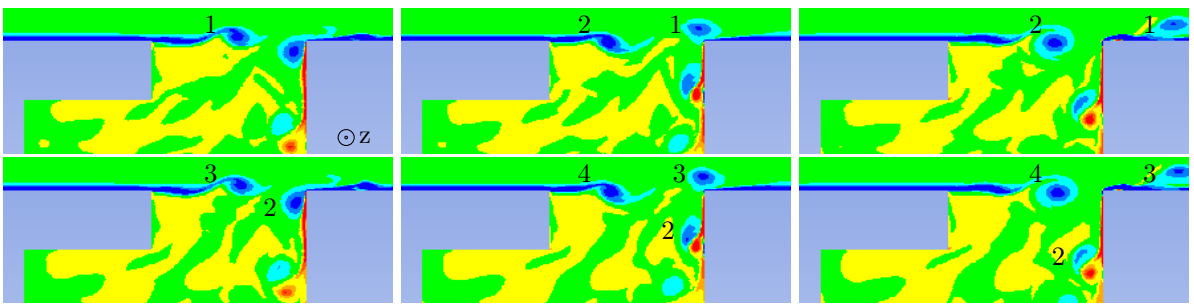
\begin{figure}[ht!]
    \centering
    \begin{tikzpicture}
            \node [inner sep=0pt, text width=0.32\textwidth]
            {\includegraphics[width=\textwidth, trim = 5.5cm 15.5cm 18cm 6cm, clip]{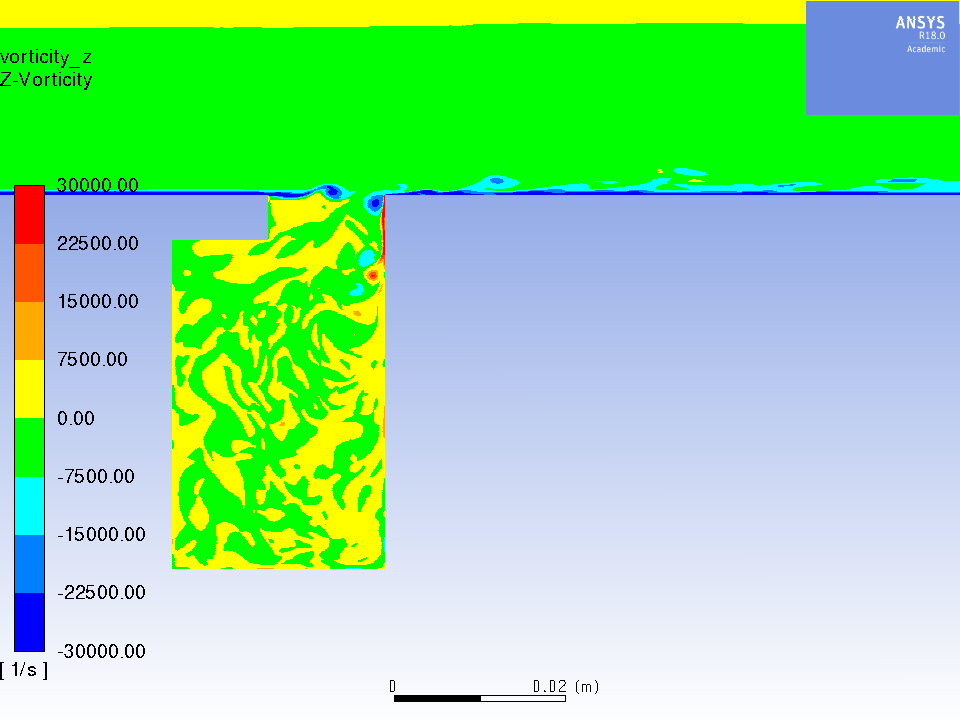}};
            \draw[draw=black, thin]  (1.95,-0.75) circle (0.5pt);
            \draw[draw=black, thin]  (1.95,-0.75) circle (3pt);
            \node (Z) at (2.20,-0.75) {z};  
            \node (1) at (0.17, 0.75) {1};  
    \end{tikzpicture}
    \begin{tikzpicture}
            \node [inner sep=0pt, text width=0.32\textwidth]
            {\includegraphics[width=\textwidth, trim = 5.5cm 15.5cm 18cm 6cm, clip]{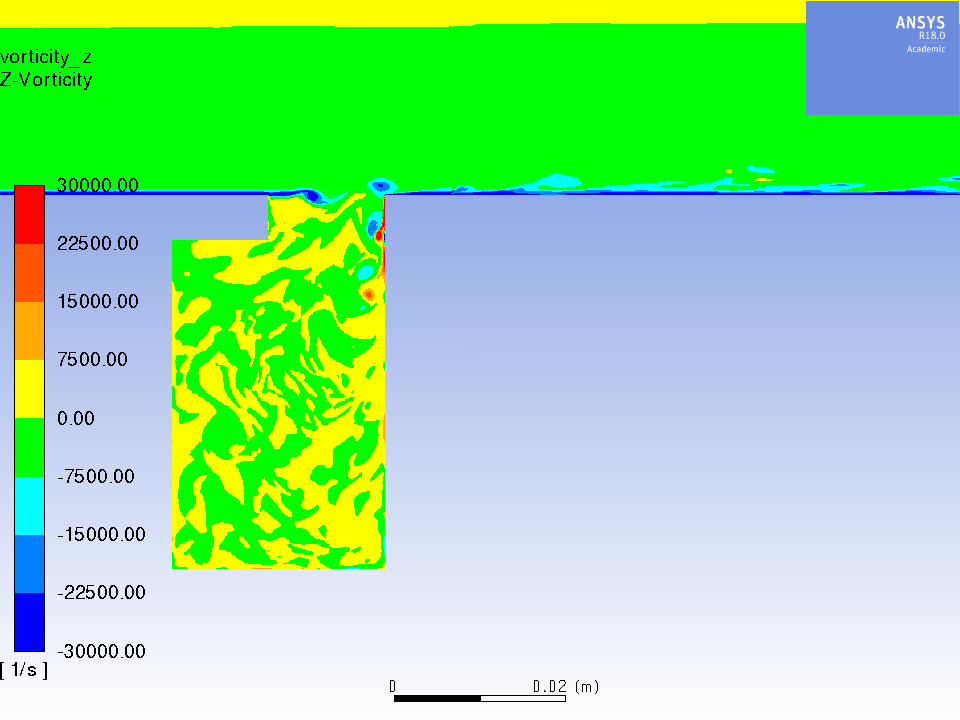}};
            \node (1) at ( 1.05, 0.75) {1};  
            \node (2) at (-0.15, 0.75) {2};  
        \end{tikzpicture}
    \begin{tikzpicture}
            \node [inner sep=0pt, text width=0.32\textwidth]
            {\includegraphics[width=\textwidth, trim = 5.5cm 15.5cm 18cm 6cm, clip]{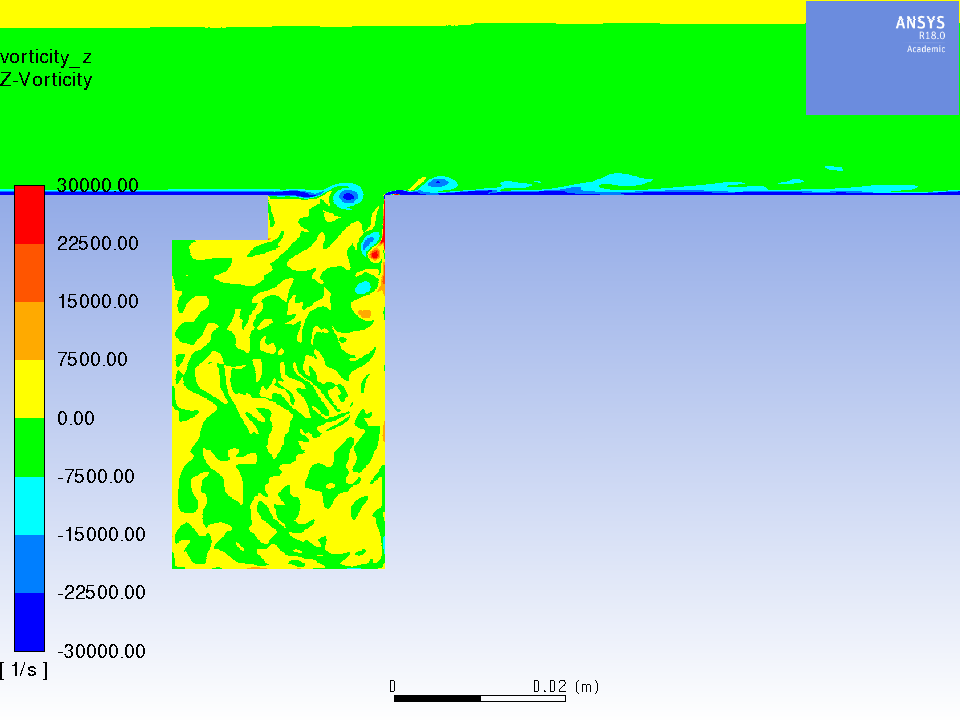}};
            \node (1) at (2.05, 0.75) {1};  
            \node (2) at (0.55, 0.75) {2};  
        \end{tikzpicture}
        \begin{tikzpicture}
            \node [inner sep=0pt, text width=0.32\textwidth]
            {\includegraphics[width=\textwidth, trim = 5.5cm 15.5cm 18cm 6cm, clip]{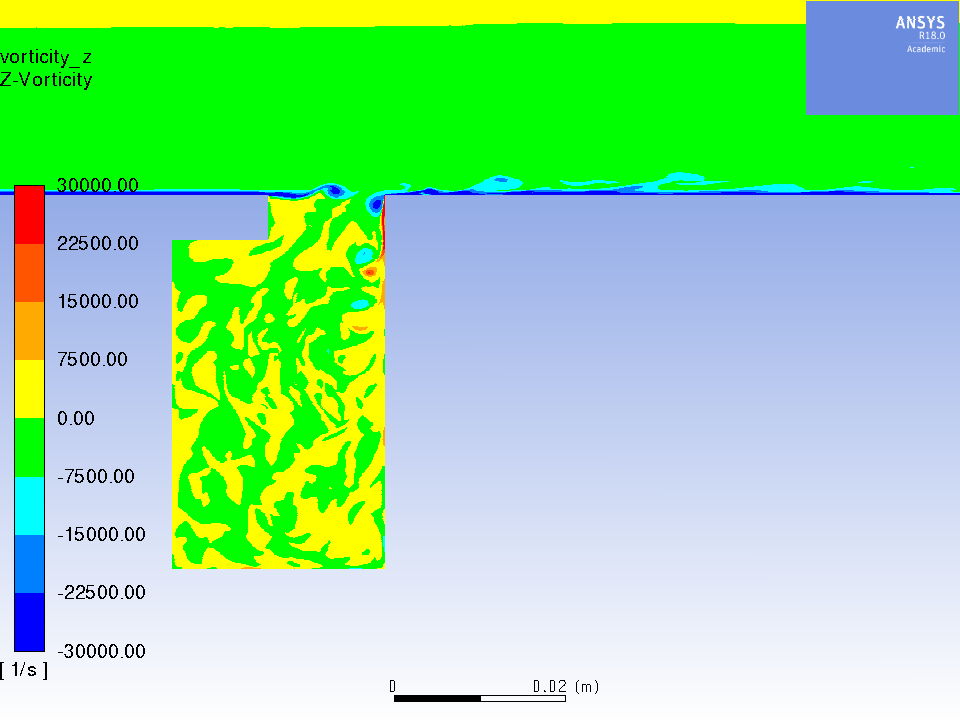}};
            \node (2) at (0.95, 0.30) {2};  
            \node (3) at (0.17, 0.75) {3};  
    \end{tikzpicture}
    \begin{tikzpicture}
            \node [inner sep=0pt, text width=0.32\textwidth]
            {\includegraphics[width=\textwidth, trim = 5.5cm 15.5cm 18cm 6cm, clip]{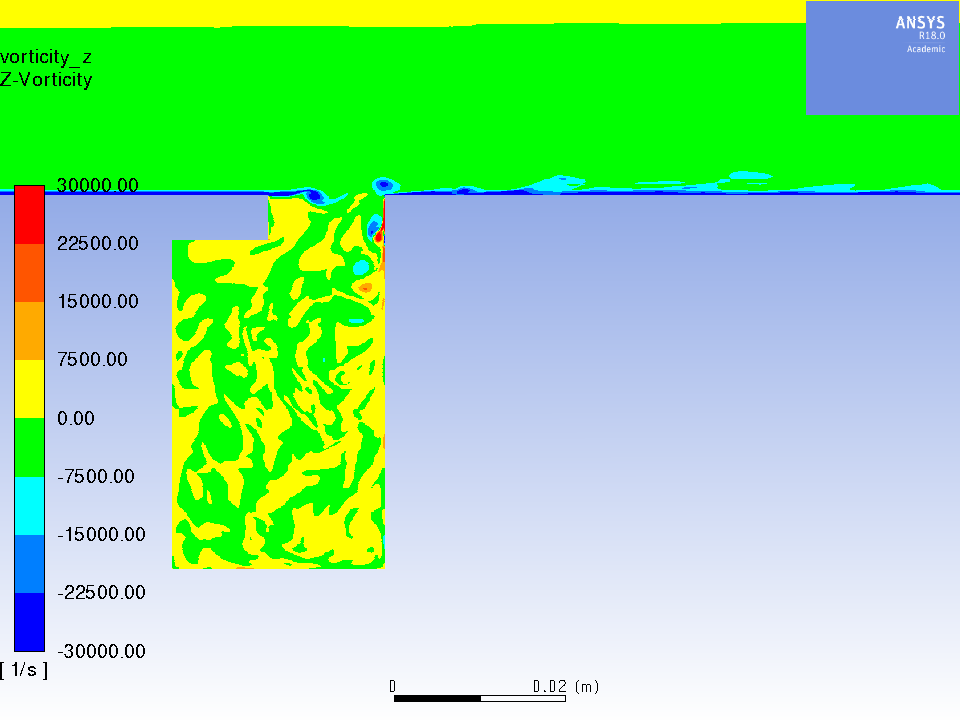}};
            \node (2) at ( 0.95, 0.00) {2};  
            \node (3) at ( 1.10, 0.75) {3};  
            \node (4) at (-0.15, 0.75) {4};  
        \end{tikzpicture}
    \begin{tikzpicture}
            \node [inner sep=0pt, text width=0.32\textwidth]
            {\includegraphics[width=\textwidth, trim = 5.5cm 15.5cm 18cm 6cm, clip]{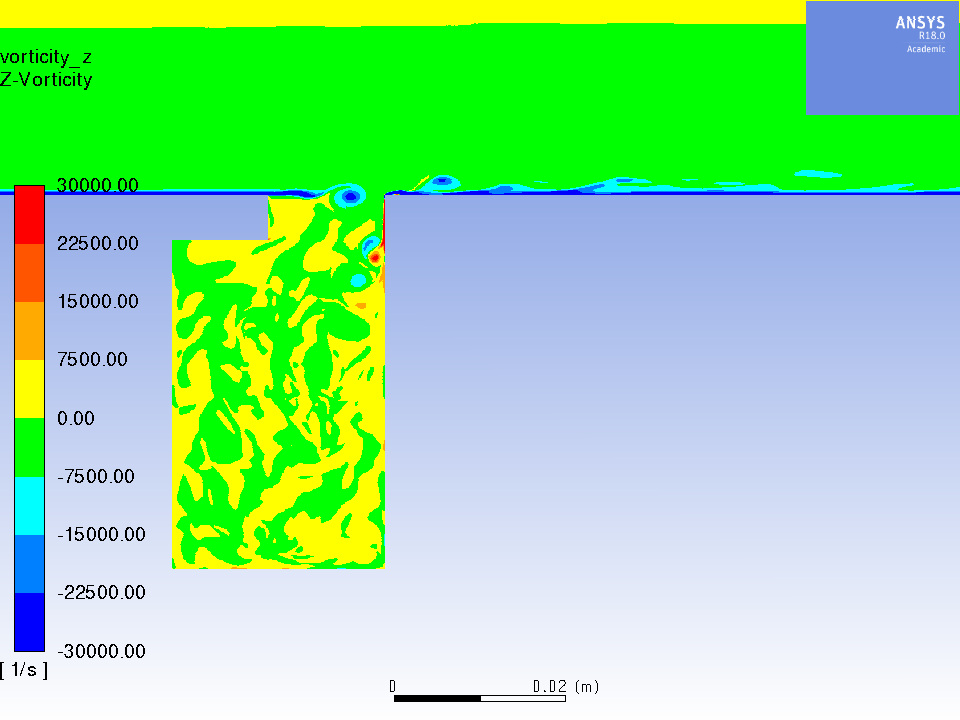}};
            \node (2) at (0.85,-0.35) {2};  
            \node (3) at (2.05, 0.75) {3};  
            \node (4) at (0.55, 0.75) {4};  
        \end{tikzpicture}
    \caption{Slice in the middle plane displaying the instantaneous field of the vorticity~$\Omega_z$ from P-U26 over one period. Snapshot sequences starts from left to the right and from top to the bottom at step: $1/126$, $22/126$, $43/126$, $64/126$, $85/126$ and $106/126$. $\Omega_z$ scales between $\unit[-3\cdot 10^{4}]{1/s}$ (blue) and $\unit[3\cdot 10^{4}]{1/s}$ (red).}
    \label{fig:zweiteRossiterMode_SPU26_Uz}
\end{figure}

After observing the flow structures in Fig.~\ref{fig:zweiteRossiterMode_SPU26_Uz}, we have found that the vertical shear layer oscillations and the vortex shedding occur at a frequency of $\unit[1600]{Hz}$. The aforementioned complete clipping interaction takes place for every second vortex with a frequency of $\unit[800]{Hz}$ (see Fig.~\ref{fig:zweiteRossiterMode_SPU26_Uz}), present as a subharmonic peak in the P-U26 spectrum (see Fig.~\ref{fig:parameterstudie_speed_FFT}).
All remaining peaks ($\unit[2411]{Hz}$, $\unit[3211]{Hz}$, $\unit[4012]{Hz}$ and $\unit[4812]{Hz}$) are higher harmonics of the shear layer oscillation.
An interesting point is that the $\unit[113.6]{dB}$ peak at $\unit[1601]{Hz}$ is as strong (in terms of pressure level) as the peak at $\unit[800]{Hz}$. We conclude that both mechanism, vortex shedding, and the interaction with the trailing edge, are energetically important effects.

\subsection{Time step size}
\begin{figure}[H]
    \centering
        \includegraphics[width=0.80\textwidth]{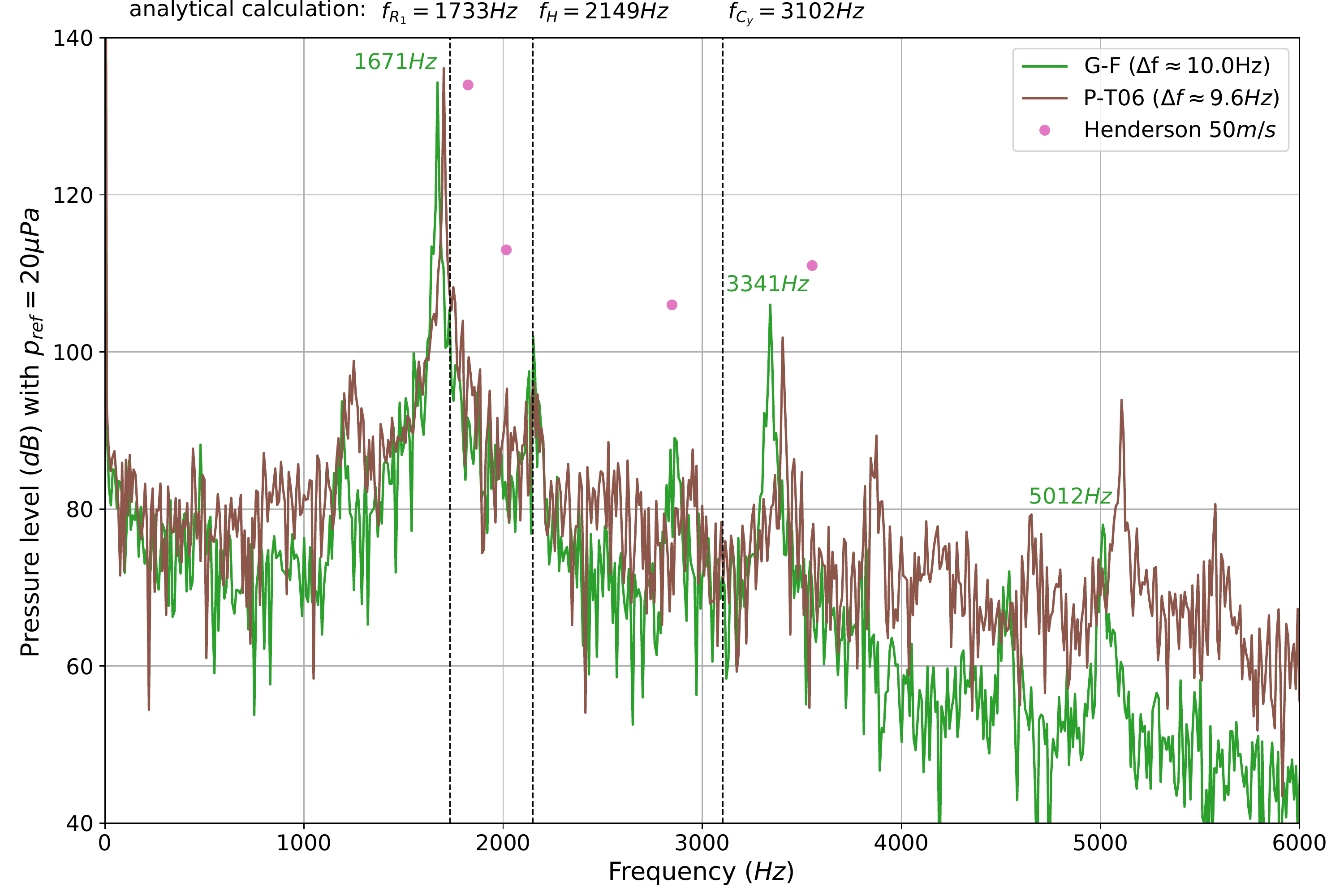} 
    \caption{Pressure level spectra for the time step size variation of the parametric study compared to the experimental data of Henderson~\cite{henderson2000}.
    }
    \label{fig:parameterstudie_time_FFT}
\end{figure}

As depicted in Fig.~\ref{fig:parameterstudie_time_FFT}, both the peak frequencies and the pressure level remain nearly unchanged for different time-step sizes. 
However, the high-frequency components are better resolved with a reduced time step size. If the flow simulation is designed for a hybrid aeroacoustic workflow, this effect on the frequency resolution must be taken into account. Nevertheless, the main flow features, up to $\unit[3500]{Hz}$, are captured well by both simulations, which justifies the use of the coarser time step size for lower frequencies. 
For a higher time resolution, we resolve additional scales between approximately $\unit[800]{Hz}$ and $\unit[1100]{Hz}$.
These scales arise due to deviations in the vortex-edge interaction, similar to the partial clipping in full escape at the low free stream velocity. According to the coherence study, incoherent structures occur at the trailing edge compared to the reference simulation.
Furthermore, the subharmonic peak at around $\unit[1250]{Hz}$, which is connected to the 3D effects driven by recirculation and vortex pairing, is more pronounced as a consequence.

\subsection{Coherence Study}
\label{sec:coherence}
\begin{figure}[h!]
    \centering
    \includegraphics[width=0.60\textwidth, trim = 0cm 0cm 0cm 0cm, clip]{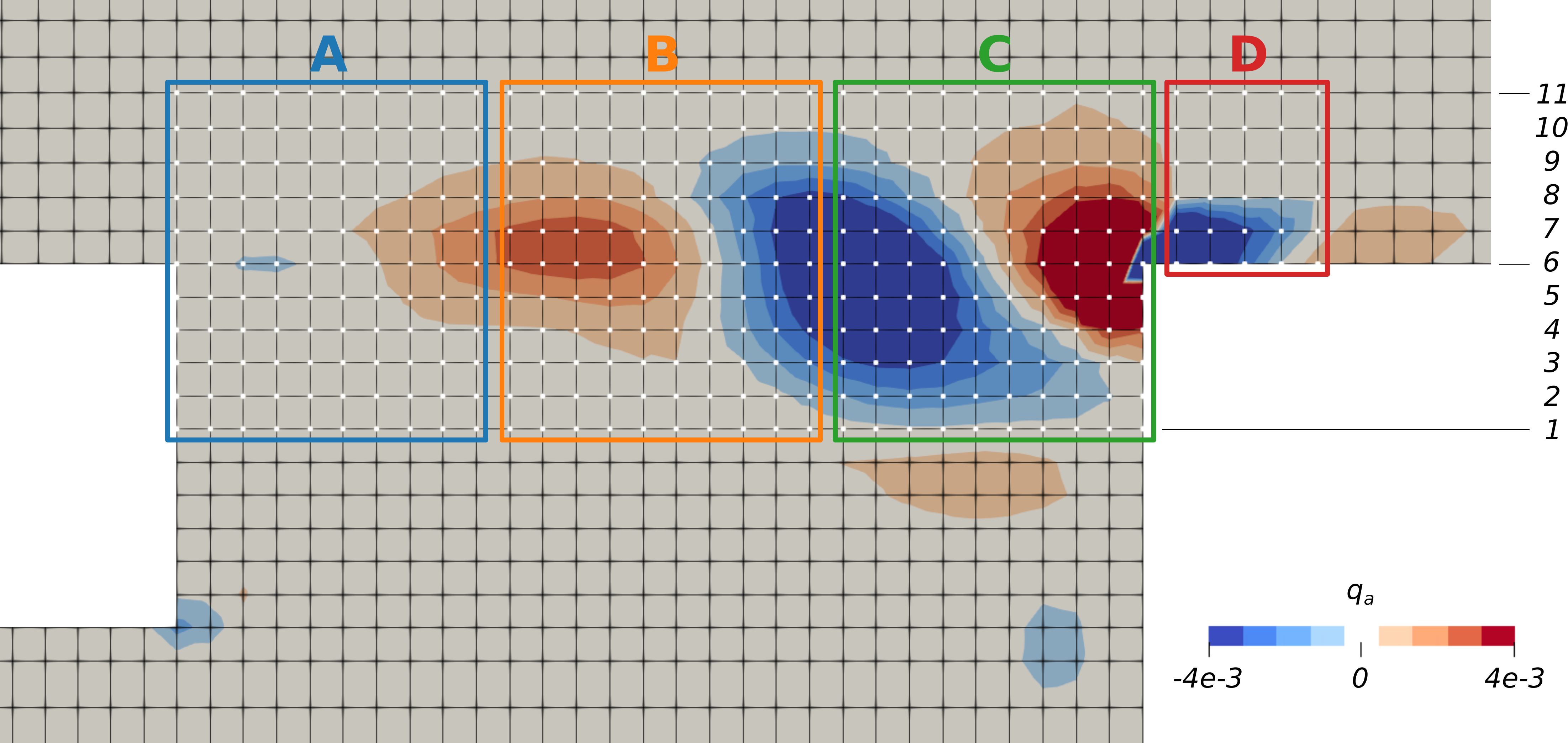}
    \caption{Segments of the coherence study's test region.}
    \label{fig:coherenceStudySamplingPositions}
\end{figure}

In this section, we verify the feasibility of reducing the acoustic computational domain from 3D to 2D for this application by conducting a coherence study of acoustically active flow structures in the spanwise direction. A well known aeroacoustic analogy is the inhomogenous wave equation of Lighthill \cite{Lighthill1952,Lighthill1954}
\begin{equation}
    \label{eq:lighthill_Wellengleichung_INHOMOGEN}
    \left( \frac{1}{\cO^2}\frac{\partial^2}{\partial t^2}    - \Delta \right)
    \left[ \cO^2 \left(\rho-\rhoO\right) \right] 
    = \div\div\tensL,
\end{equation}
with the speed of sound~$c$, the density~$\rho$ of the real fluid, the density~$\rhoO$ of an ideal linear acoustic fluid, and the Lighthill stress tensor~$\tensL$.
Due to the low Mach number ($M^2<<1$) and after neglecting thermal and dissipative viscous effects we approximate the Lighthill stress tensor as
\begin{equation}
    \label{eq:LH_stressTensor}
    \tensL\approx\rho\vecu\vecu.
\end{equation}
Using Equ.~\eqref{eq:LH_stressTensor}, we performed a comprehensive coherence study for~$q_\textrm{a} = \div\div\tensL$, which allowed us to investigate the source field properties in the spanwise dimension.
For these purposes, we used 54 equidistant acoustic source term probes $q_{\textrm{a,}i}$ in the spanwise direction in the region near the cavity mouth, where the dominant sources occur (see Fig.~\ref{fig:coherenceStudySamplingPositions}).
After defining a reference probe $q_{\textrm{a,ref}}$ at the middle of the cavity's span, the coherence 
\begin{equation}
    \coh = \frac{|\csd|^2}{ \psdx \cdot \psdy},  ~~~~~~~~~  0 \leq \coh \leq 1
    \label{eq:coherenceDefinition}
\end{equation}
was calculated with regard to all other 53 probes.
In Equ.~\eqref{eq:coherenceDefinition} $\csd$ denotes the cross spectral density between reference probe and probe $i$, whereas $\psdx$ and $\psdy$ denote the power spectral densities of both probes, $q_{\textrm{a,ref}}$ and $q_{\textrm{a,}i}$~\cite{bendat1980}.

The results of the coherence study for G-F and P-T06 are summarized in Tab.~\ref{tab:srcCoherenceLH}.
To distinct the coherence over the free shear layer growth, we subdivided the region of interest into four segments, denoted with A, B, C, and D as displayed in Fig.~\ref{fig:coherenceStudySamplingPositions}.
It should be noted that the coherence results are presented for a frequency of $\unit[1671]{Hz}$, which was the dominant frequency.
Inside the cavity neck, 3D effects are dominant, whereas in plane 6 and 11 the Lighthill sources cohere and behave two-dimensional.
In the case of P-T06, we observe 3D effects in plane 11 of segment D that results from the impingement process, which occurs (in contrast to G-F) due to the temporary finer resolved vortex structures.

According to Larchev\^eque~\cite{larcheveque2007}, the use of side walls generates a bifurcated flow that often leads to a switch in the dominant Rossiter mode and non-linear energy transfer.
Nevertheless, our coherence study showed that in our case the cavity sidewalls pose minor influence on the flow structures in terms of 3D effects (see Tab.~\ref{tab:srcCoherenceLH}).

\begin{table}[ht!]
  \begin{center}
      \begin{tabular}{cc}
          \toprule 
        \textbf{G-F} & \textbf{P-T06}\\
          \midrule
        \includegraphics[width=0.40\textwidth, trim = 0cm 2cm 0cm 0cm, clip]{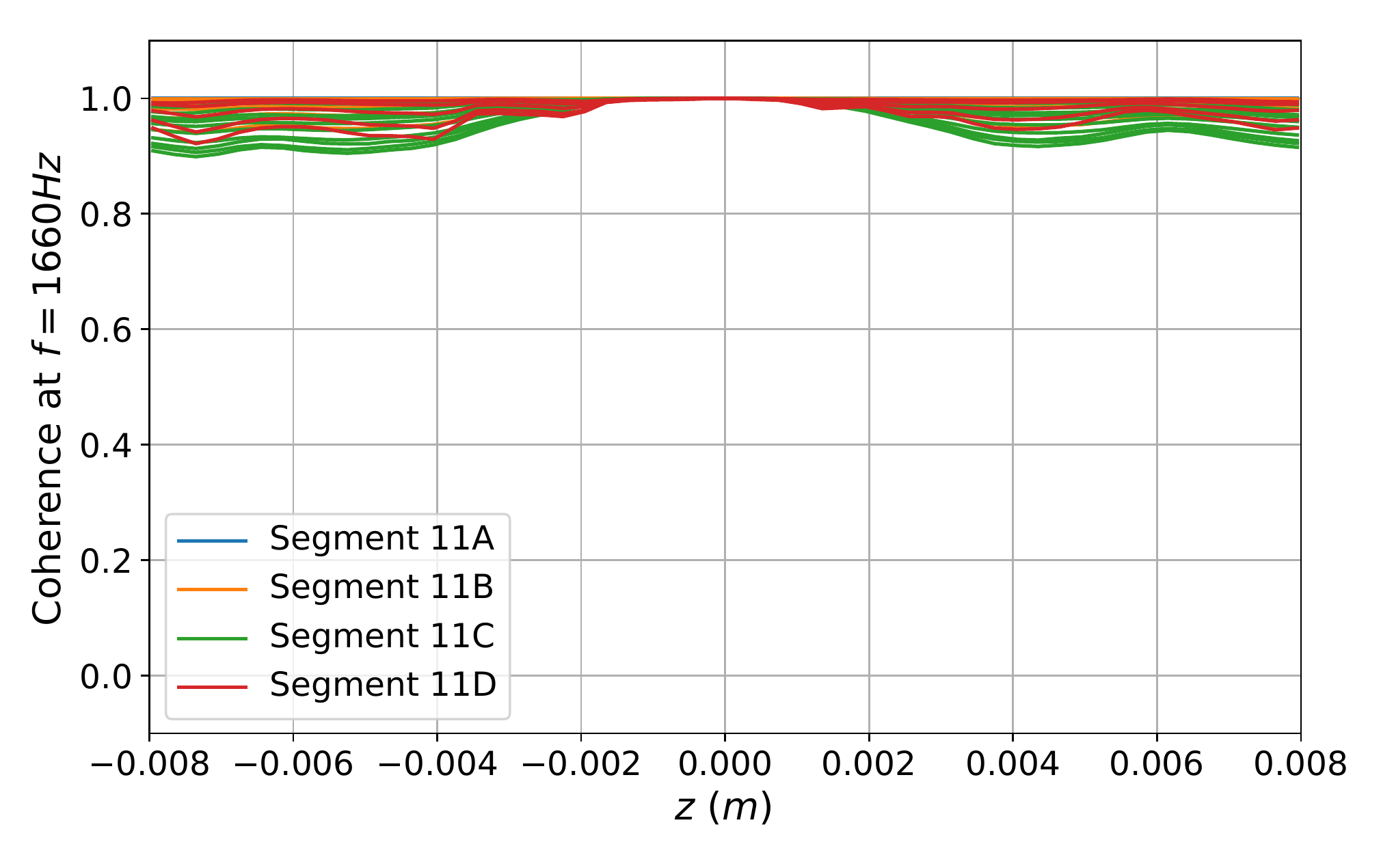} &
        \includegraphics[width=0.40\textwidth, trim = 0cm 2cm 0cm 0cm, clip]{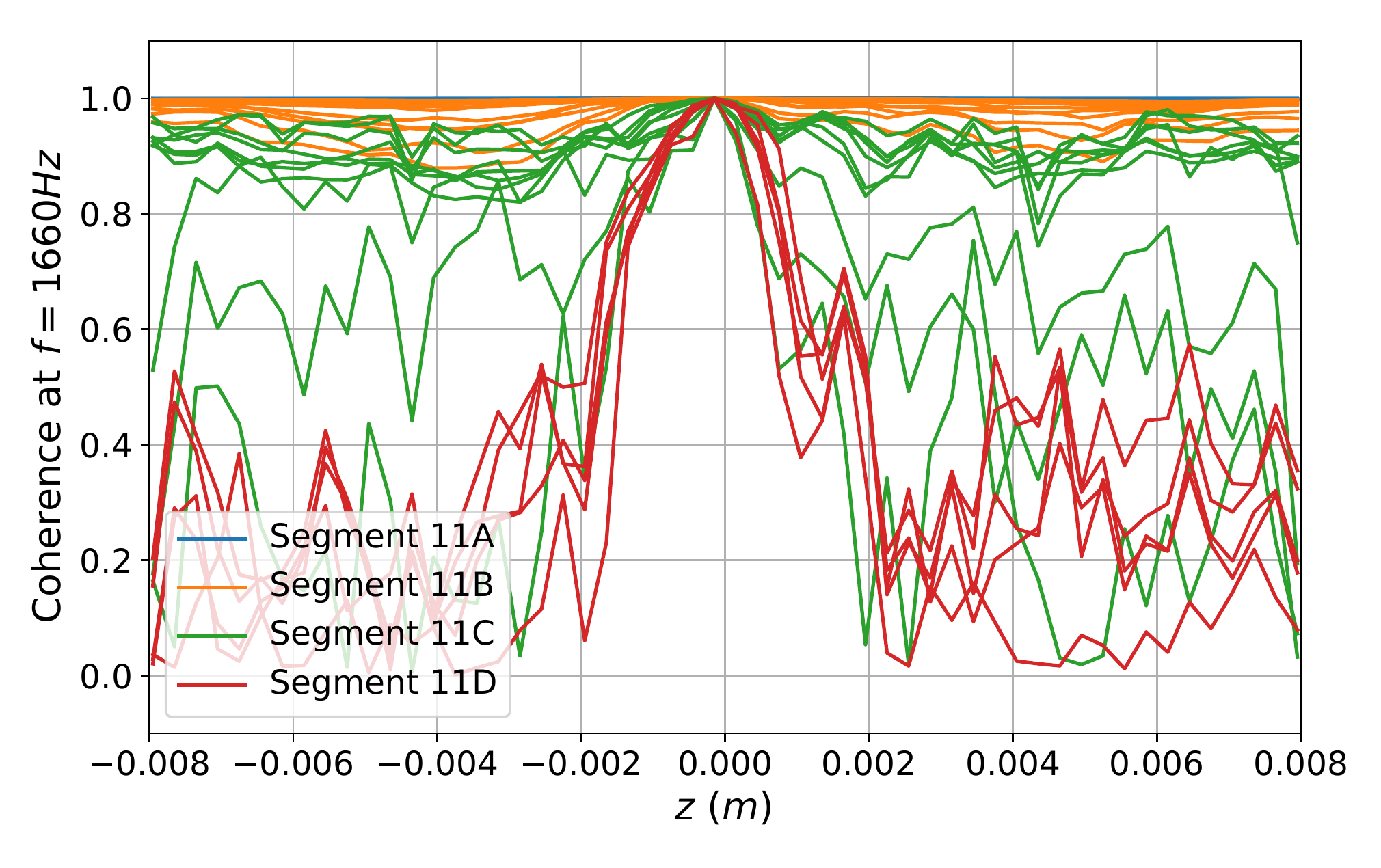} \\

        \includegraphics[width=0.40\textwidth, trim = 0cm 2cm 0cm 0cm, clip]{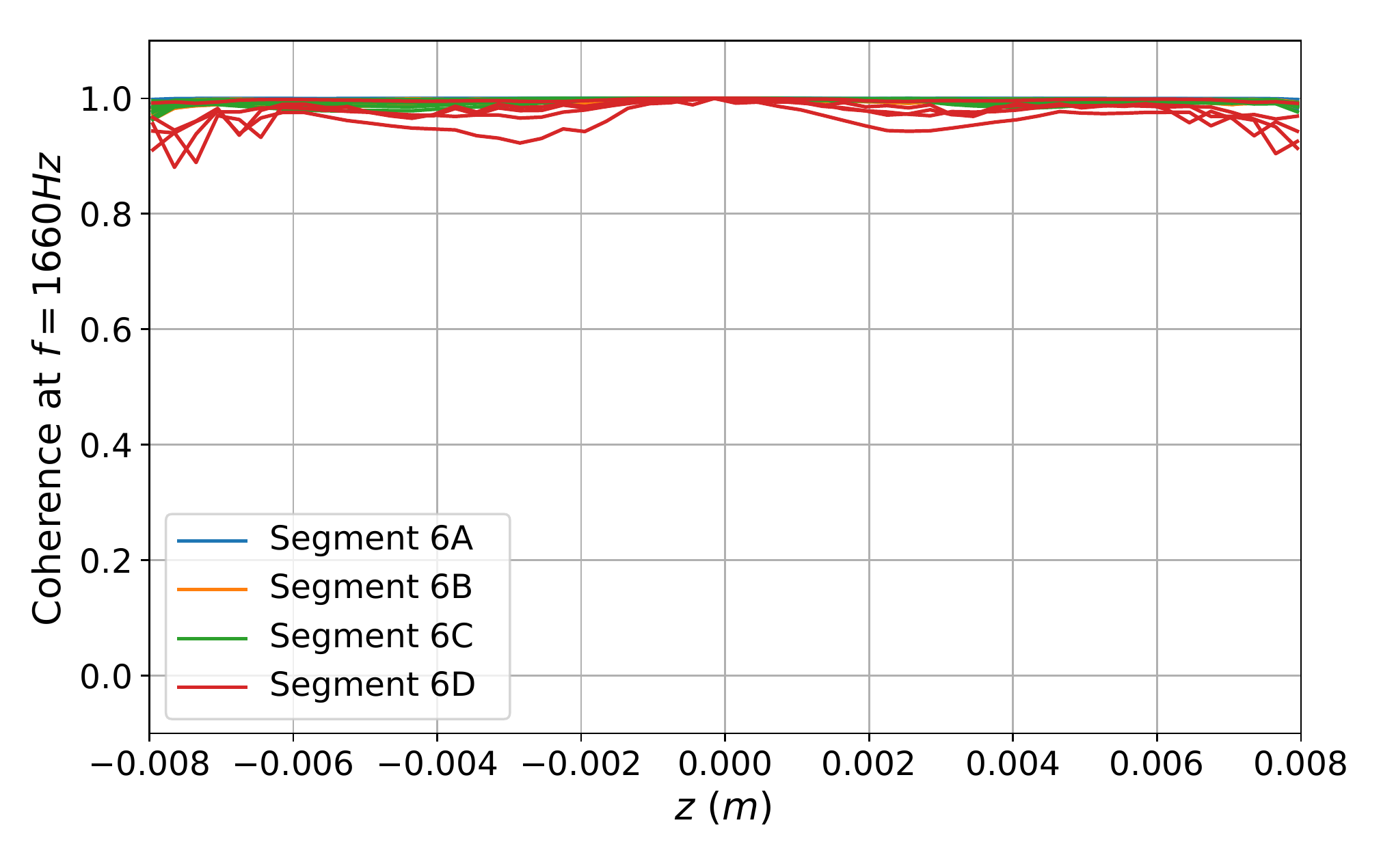} &
        \includegraphics[width=0.40\textwidth, trim = 0cm 2cm 0cm 0cm, clip]{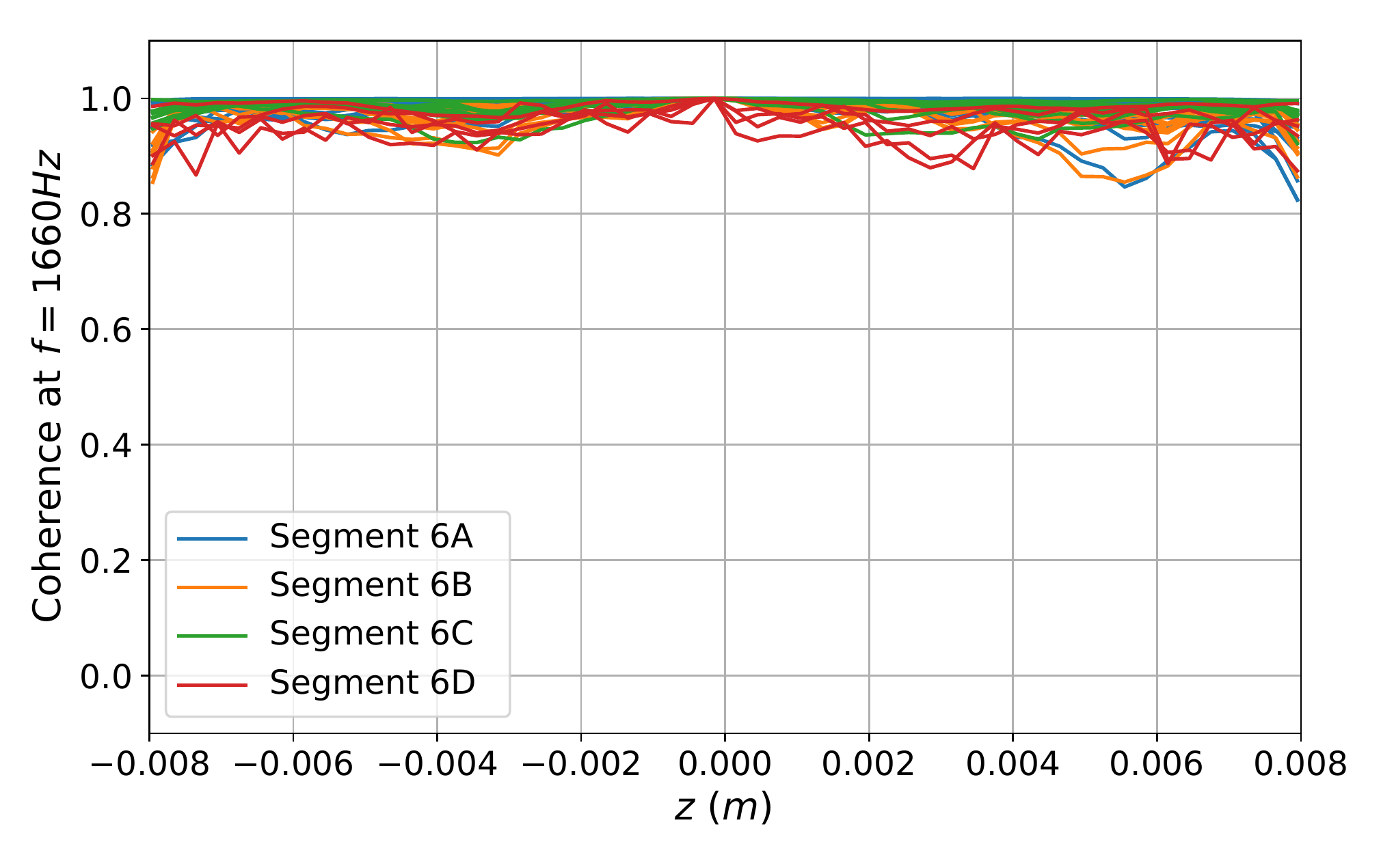} \\

        \includegraphics[width=0.40\textwidth, trim = 0cm 0cm 0cm 0cm, clip]{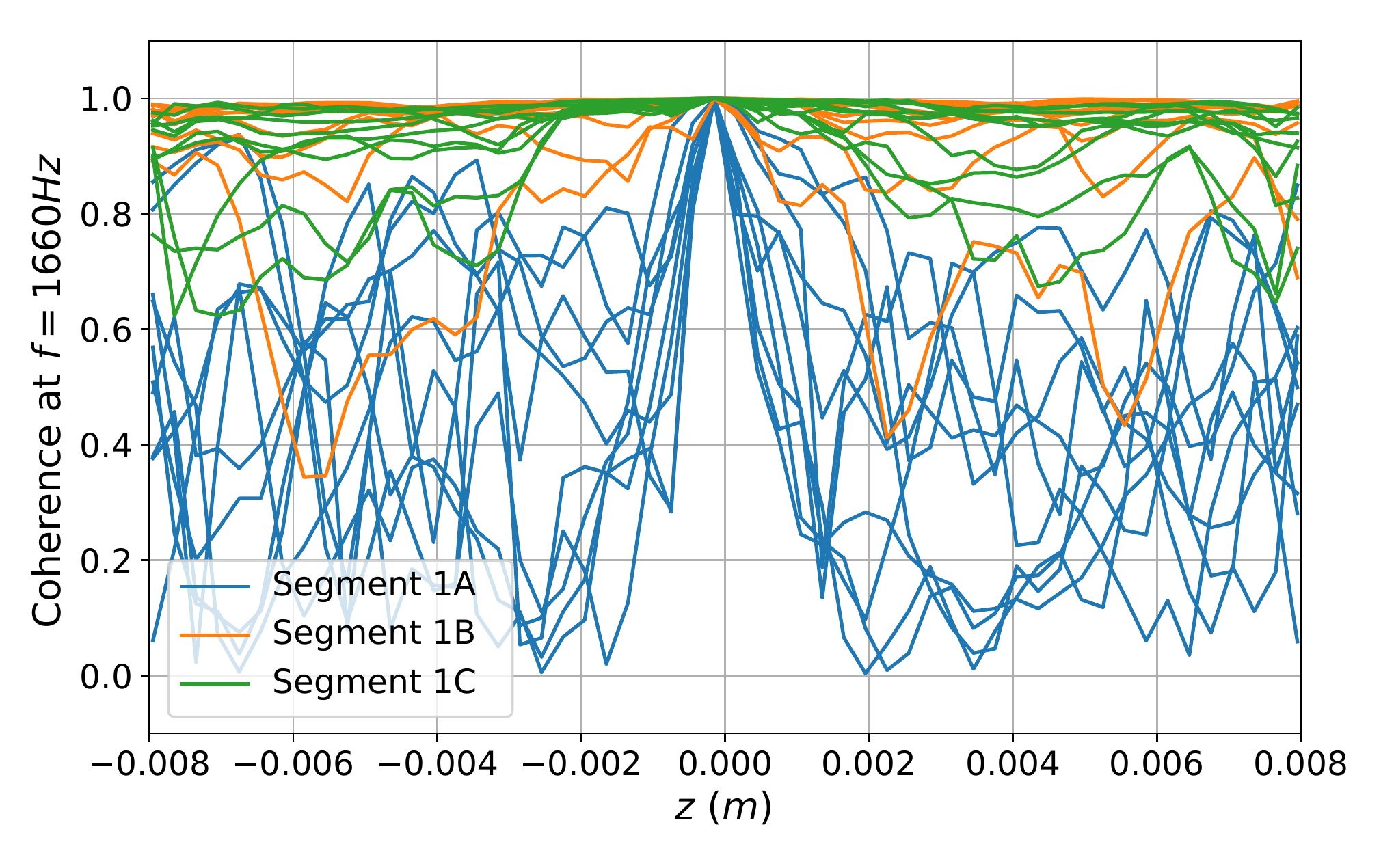} &
        \includegraphics[width=0.40\textwidth, trim = 0cm 0cm 0cm 0cm, clip]{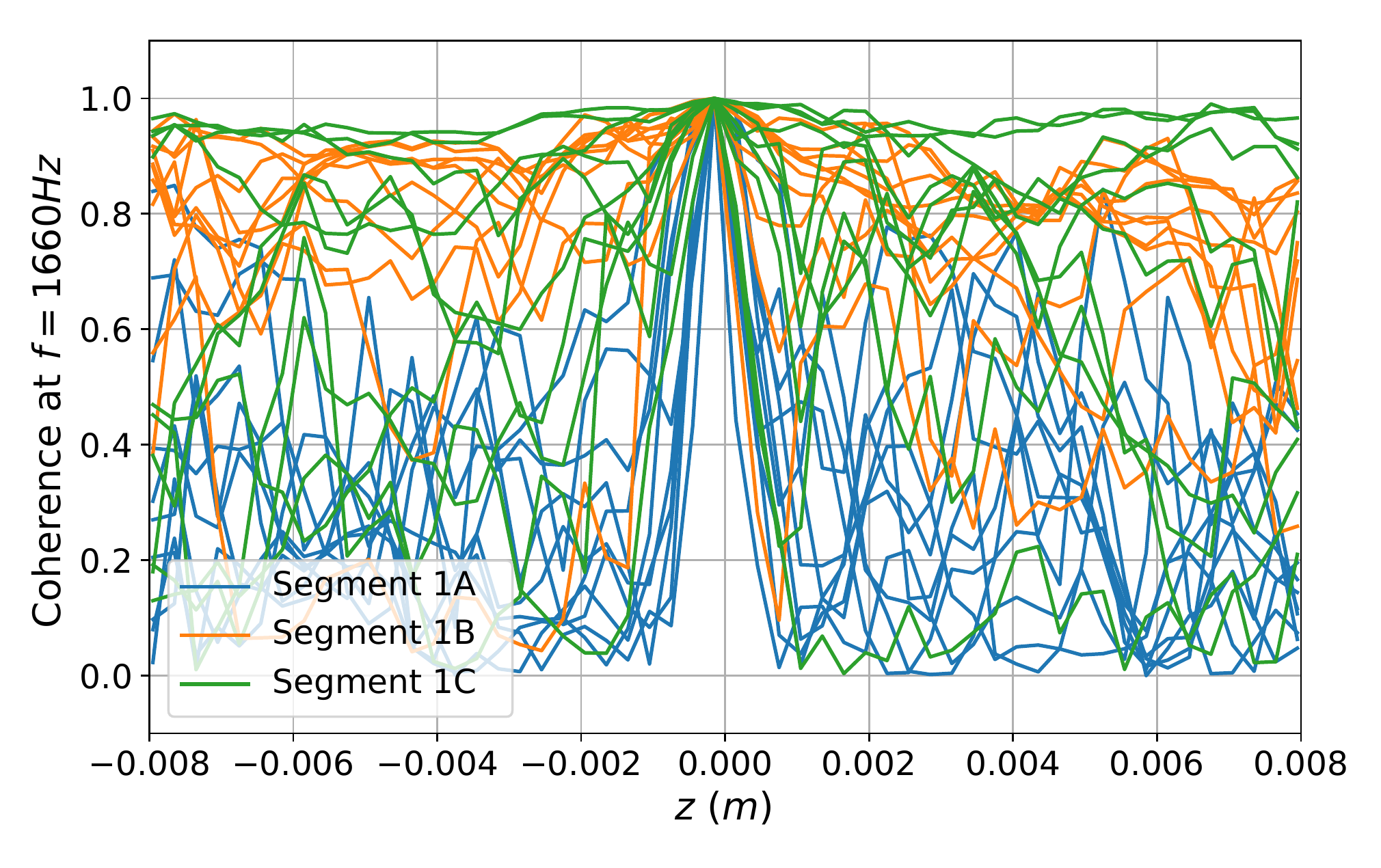} \\
        \bottomrule
      \end{tabular}
      \caption{Coherence of the Lighthill source terms from G-F and P-T06 in the span wise direction for the $1^{\mathrm{st}}$, $6^{\mathrm{th}}$ and $11^{\mathrm{th}}$ planes of the investigated region. Note that the coherence is plotted for a frequency of $\unit[1660]{Hz}$.}
      \label{tab:srcCoherenceLH}
  \end{center}
\end{table}

\section{Discussion}
\label{sec:fazitUNDausblick}
According to the third CAA workshop~\cite{nasa3CAABenchmark2000} two edge tone frequencies are expected between $\unit[0]{Hz}$ and $\unit[2000]{Hz}$, whereas frequencies related to the longitudinal cavity modes occur between $\unit[2000]{Hz}$ and $\unit[4000]{Hz}$. Table~\ref{tab:ergebnisse_f_SPL_zusammenfassung} summarizes and compares the dominant peaks from the simulation cases G-F and P-T06 to the experimental data in~\cite{henderson2000}. We associate each peak with a source mechanism and discuss it accordingly.
This simulation captures the $1^{\mathrm{st}}$ Rossiter mode and its higher harmonics ($\unit[3341]{Hz}$ and $\unit[5012]{Hz}$) at the microphone position. The amplitude of the $1^{\mathrm{st}}$ Rossiter mode ($\unit[134.3]{dB}$) is well reproduced compared to the experimental data of Henderson~\cite{henderson2000} ($\unit[134]{dB}$).
Although the peak frequency is underestimated, this discrepancy could be explained by the differences in the boundary layer thickness used in our simulation and those presented in the measurements. As shown in the parametric study, a boundary layer thickness of about $\delta = \unit[8.06]{mm}$ matches the measured peak frequency.
For a higher time resolution, we resolve additional scales between approximately $\unit[800]{Hz}$ and $\unit[1100]{Hz}$. This is due to deviations in the vortex-edge interaction, similar to the partial clipping and full escape at the low free stream velocity.

We assign the peak at $\unit[2151]{Hz}$ to the expected Helmholtz resonance ($f_\mathrm{H}=\unit[2149]{Hz}$), close to the results of other studies~\cite{henderson2000,loh2004}.
We explain the origin of the peak around ($\unit[89]{dB}$, $\unit[2861]{Hz}$) by a transversal cavity mode in depth direction (analytically at $f_\mathrm{C_y}=\unit[3102]{Hz}$) modulated by the cavity orifice.

The peak at $\unit[1190]{Hz}$ in the pressure spectrum of the simulation may be a result of recirculation or vortex pairing. As already mentioned during the discussion of Fig.~\ref{fig:SRC_coherence_explanation}, a 3D vortex below the leading edge convects vertically, participates directly in the new vortex formation, and influences the shear layer growth instability. Nevertheless, further studies are needed to classify the origin of this peak. 

\begin{table}[ht!]
 \begin{center}
    \begin{tabular}{cc}
        \multicolumn{2}{c}{\textbf{Henderson}}\\
        \toprule
          $\boldsymbol{f}$ & $\boldsymbol{PL}$\\
          $(\unit{Hz})$ & $(\unit{dB})$ \\
          \midrule
          $-$  & $-$\\
          $930$  & $103$\\
          $1340$ & $107$\\
          $1824$ & $134$\\
          $2016$ & $113$\\
          $2848$ & $106$\\
          $3552$ & $111$\\
          $-$ & $-$\\
          $-$ & $-$\\
          $-$ & $-$\\
          $-$ & $-$\\
          \bottomrule
    \end{tabular}
    \hspace{1mm}
    \begin{tabular}{ccccc}
        \multicolumn{2}{c}{\textbf{G-F}} & \multicolumn{2}{c}{\textbf{P-T06}} &  \\
        \toprule
          $\boldsymbol{f}$ & $\boldsymbol{PL}$ & $\boldsymbol{f}$ & $\boldsymbol{PL}$ &  \multirow{2}{*}{\textbf{Mechanism}}\\
          $(\unit{Hz})$ & $(\unit{dB})$  & $(\unit{Hz})$ & $(\unit{dB})$  &  \\
          \midrule
          $480$  & $88.2$  & $442$  & $87.7$  & artificial domain resonance \\
          $-$    & $-$     & $798$  & $87.1$  & shear layer-edge interaction \\ 
          $1190$ & $93.7$  & $1250$ & $98.9$  &  3D effects$^{*}$\\
          $1671$ & $134.3$ & $1702$ & $136.1$ & $1^{\mathrm{st}}$ Rossiter mode $f_\mathrm{R}$ \\
          $2152$ & $102$   & $2154$ & $96.2$  & Helmholtz resonance $f_\mathrm{H}$\\
          $2861$ & $89$    & $2952$ & $87.6$  & transversal cavity mode $f_\mathrm{C_y}$\\
          $3341$ & $106$   & $3404$ & $101.8$ & $1^{\mathrm{st}}$ harmonic of $f_\mathrm{R}$\\
          $3822$ & $75.7$  & $3875$ & $89.4$  & unknown \\
          $4543$ & $72$    & $4654$ & $79.3$  & $1^{\mathrm{st}}$ harmonic$^{*}$ of $f_\mathrm{H}$ \\
          $5012$ & $78$    & $5106$ & $93.9$  & $2^{\mathrm{nd}}$ harmonic of $f_\mathrm{R}$\\
          $5502$ & $58.3$  & $5577$ & $80.6$  & $1^{\mathrm{st}}$ harmonic$^{*}$ of $f_\mathrm{C_y}$ \\
          \bottomrule
    \end{tabular}
     \caption{Numerical pressure peak values at the high flow velocity ($\unit[50]{m/s}$) compared to the experimental data of Henderson~\cite{henderson2000}. Mechanisms denoted with $\Box^{*}$ are based on assumptions and further investigations are needed.}
     \label{tab:ergebnisse_f_SPL_zusammenfassung}
 \end{center}
\end{table}

\section{Conclusion}
\label{sec:end}
In this paper, we investigated a generic deep cavity with an overhanging lip that is overflowed by air at two different free stream velocities, namely $\unit[26.8]{m/s}$ and $\unit[50]{m/s}$. This cavity geometry and the problem definition were initially introduced and experimentally studied within the \textit{Third Computational Aeroacoustics (CAA) Workshop on Benchmark Problems} by NASA~\cite{nasa3CAABenchmark2000} and Henderson~\cite{henderson2000}, respectively. The turbulent boundary layer and the acoustic waves interact with the cavity's geometry and form a strong feedback mechanism.

%
%
The present work focused on the details of the compressible turbulent flow structures and their variations concerning the velocity, the boundary layer as well as the domain dimensionality for a later acoustic simulation within a hybrid aeroacoustic workflow.
For these purposes, we resolved the large turbulent scales by the DES-based SBES turbulence model on a 3D domain.
In the far-field, we treated the acoustic component with acoustically absorbing boundaries based on the radiation characteristics and succeeded to reduce the reflections from the domain boundaries, which are proven to contaminate the results (see~\cite{kurbatskii2000,ashcroft2003,farkasANDpaal2015}).
Furthermore, we quantified the most appropriate domain size and grid density for the numerical simulation within a grid study.

In contrast to URANS turbulence models, where only the $1^{\mathrm{st}}$ Rossiter mode and its higher harmonics are captured, the SBES model resolved more turbulent structures and therefore acoustic resonant effects.
During the grid and parameter studies, we assessed the broadband pressure spectrum at the evaluation position, where the peak frequencies and pressure levels are inversely dependent on the thickness of the approaching boundary layer above the leading edge of the cavity.
Close to the results of other studies~\cite{henderson2000,loh2004}, we assigned the peak at $\unit[2151]{Hz}$ to the expected Helmholtz resonance ($f_\mathrm{H}=\unit[2149]{Hz}$).
Although Henderson~\cite{henderson2000} had no explanation for the peak around $\unit[2861]{Hz}$, we assumed this acoustic resonance to be a transversal cavity duct mode in the depth direction (analytically at $f_\mathrm{C_y}=\unit[3102]{Hz}$) modulated by the complicated cavity orifice.
Additionally to the physical peaks, we detected an artificial computational domain resonance at $\unit[480]{Hz}$ of $\unit[88.2]{dB}$ in the pressure spectrum of the G-F simulation.
This non-physical resonance arises due to not fully absorbing boundary conditions but has a small influence on the computations.
For the coarse and medium grid density, this peak is masked by the turbulent fluctuations at the measurement location.

Furthermore, we verified the feasibility of reducing the acoustic computational domain from 3D to 2D for this application by conducting a coherence study of acoustically active flow structures in the spanwise direction.
This coherence study showed that in our case the cavity side walls have a minor influence on the acoustically active flow structures in terms of 3D effects.
After visualizing relevant flow structures, we investigated and determined the vortex-edge interactions as previously observed in experiments (see Rockwell and Knisely~\cite{rockwellANDknisely80}).
Similar to Ashcroft et al.~\cite{ashcroft2003}, the 3D Taylor-G{\"o}rtler vortices glide along the lower edge of the cavity lip thus creating a vertical flow that is convected towards the shear layer (see Fig.~\ref{fig:SRC_coherence_explanation}).
Remarkably for the lower approaching velocity ($\unit[26.8]{mm}$), we found a special vortex-edge interaction, namely an alternating sequence of complete clipping and a subsequent partial escape.

\section*{Acknowledgements}
\label{sec:Acknowledgements}
The CFD simulations performed for this study were achieved in part using the Vienna Scientific Cluster.
The ANSYS Fluent licenses were kindly provided by ANSYS Germany GmbH.

The high-performance computing (HPC) time needed for this study was kindly provided by \textit{Vienna Scientific Cluster} (\textit{VSC}).


\section*{References}
\bibliographystyle{elsarticle-num}

\begin{thebibliography}{10}
\expandafter\ifx\csname url\endcsname\relax
  \def\url#1{\texttt{#1}}\fi
\expandafter\ifx\csname urlprefix\endcsname\relax\def\urlprefix{URL }\fi
\expandafter\ifx\csname href\endcsname\relax
  \def\href#1#2{#2} \def\path#1{#1}\fi

\bibitem{chu58}
B.~T. Chu, L.~S.~G. Kovasznay, Non-linear interactions in a viscous
  heat-conducting compressible gas.

\bibitem{rockwell78}
D.~Rockwell, E.~Naudasher, Review -- self-sustaining oscillations of flow past
  cavities.

\bibitem{east66}
L.~F. East, Aerodynamically induced resonance in rectangular cavities.

\bibitem{nasa3CAABenchmark2000}
Third computational aeroacoustics (caa) workshop on benchmark problems --
  category 6.

\bibitem{henderson2000}
B.~Henderson, Automobile noise involving feedback -- sound generation by low
  speed cavity flows.

\bibitem{farkasANDpaal2015}
B.~Farkas, G.~Paal, Numerical study on the flow over a simplified vehicle door
  gap -- an old benchmark problem is revisited.

\bibitem{ahujaUndMendoza95}
K.~K. Ahuja, J.~Mendoza, Effects of cavity dimensions, boundary layer, and
  temperature on cavity noise with emphasis on benchmark data to validate
  computational aeroacoustic codes.

\bibitem{lazarov2018}
I.~Lazarov, Aeroacoustic simulation of a deep cavity with a lip, master thesis,
  Master's thesis (Apr. 2018).

\bibitem{moon2000}
Y.~J. Moon, S.~R. Koh, Y.~Cho, J.~M. Chung, Aeroacoustic computations of the
  unsteady flows over a rectangular cavity with lip.

\bibitem{kurbatskii2000}
K.~K. Kurbatskii, C.~K.~W. Tam, Direct numerical simulation of automobile
  cavity tones.

\bibitem{ashcroft2000}
G.~B. Ashcroft, K.~Takeda, X.~Zhang, Computations of self-induced oscillatory
  flow in an automobile door cavity.

\bibitem{lin2004}
W.~H. Lin, R.~H. Loh, Numerical solutions to the fourth and second
  computational aeroacoustics (caa) workshop benchmark problems.

\bibitem{zhang2004}
Z.~Zhang, R.~Barron, C.-F. An, Spectral analysis for air flow over a cavity.

\bibitem{loh2004}
C.~Y. Loh, P.~C.~E. Jorgenson, Computation of tone noises generated in viscous
  flows.

\bibitem{nasa4CAABenchmark2004}
Fourth computational aeroacoustics (caa) workshop on benchmark problems --
  category 5, problem 2.

\bibitem{ashcroft2003}
G.~B. Ashcroft, K.~Takeda, X.~Zhang, A numerical investigation of the noise
  radiated by a turbulent flow over a cavity.

\bibitem{koh2003}
S.~R. Koh, Y.~Cho, Y.~J. Moon, Aeroacoustic computation of cavity flow in
  self-sustained oscillations.

\bibitem{wang2007}
Z.~K. Wang, G.~Djambazov, C.~H. Lai, K.~Pericleous, Numerical simulation of
  flow-induced cavity noise in self-sustained oscillations.

\bibitem{ansysTheory}
ANSYS Fluent Theory Guide (Release 15.0) (2013).

\bibitem{rockwellANDknisely80}
D.~Rockwell, C.~Knisely, Vortex-edge interaction: Mechanisms for generating low
  frequency components.

\bibitem{Lighthill1952}
M.~J. Lighthill, {On sound generated aerodynamically I. General theory},
  Proceedings of the Royal Society of London 211 (1951) 564--587.

\bibitem{Lighthill1954}
M.~J. Lighthill, {On sound generated aerodynamically II. Turbulence as a source
  of sound}, Proceedings of the Royal Society of London 222 (1953) 1--32.

\bibitem{bendat1980}
J.~S. Bendat, A.~G. Piersol, Engineering applications of correlation and
  spectral analysis.

\bibitem{larcheveque2007}
L.~Larchev\^eque, P.~Sagaut, O.~Labb\'e, Large-eddy simulation of a subsonic
  flow including asymmetric three-dimensional effects.

\end{thebibliography}

\end{document}